\documentclass[ALICE,manyauthors]{cernphprep}
\usepackage[comma,square,numbers,sort&compress]{natbib}
\usepackage{hyperref}
\usepackage{lineno}
\usepackage{xspace}
\usepackage{color}
\usepackage{placeins}
\usepackage{subfigure}

\usepackage[document]{ragged2e}
\usepackage[T1]{fontenc}
\usepackage{orcidlink}

\begin{document}
%
\newcommand{\Kzs}{\rm K^{ 0}_S}
\newcommand{\Kpm}{\rm K^{\pm}}
\newcommand{\Kz}{\rm K^0}
\newcommand{\Kzb}{\rm \overline{K}\,^0}
\newcommand{\Kplus}{\rm K^+}
\newcommand{\KpKm}{\rm K^+ K^-}
\newcommand{\Kmin}{\rm K^-}
\newcommand{\Kaon}{\rm K}
\newcommand{\fz}{\rm f_{\rm 0}}
\newcommand{\az}{\rm a_{\rm 0}}
\newcommand{\pro}{\rm p}

\newcommand{\pp}           {pp\xspace}
\newcommand{\ppbar}        {\mbox{$\mathrm {p\overline{p}}$}\xspace}
\newcommand{\XeXe}         {\mbox{Xe--Xe}\xspace}
\newcommand{\PbPb}         {\mbox{Pb--Pb}\xspace}
\newcommand{\pA}           {\mbox{pA}\xspace}
\newcommand{\pPb}          {\mbox{p--Pb}\xspace}
\newcommand{\AuAu}         {\mbox{Au--Au}\xspace}
\newcommand{\dAu}          {\mbox{d--Au}\xspace}

\newcommand{\s}            {\ensuremath{\sqrt{s}}\xspace}
\newcommand{\snn}          {\ensuremath{\sqrt{s_{\mathrm{NN}}}}\xspace}
\newcommand{\pt}           {\ensuremath{p_{\rm T}}\xspace}
\newcommand{\meanpt}       {$\langle p_{\mathrm{T}}\rangle$\xspace}
\newcommand{\ycms}         {\ensuremath{y_{\rm CMS}}\xspace}
\newcommand{\ylab}         {\ensuremath{y_{\rm lab}}\xspace}
\newcommand{\etarange}[1]  {\mbox{$\left | \eta \right |~<~#1$}}
\newcommand{\yrange}[1]    {\mbox{$\left | y \right |~<~#1$}}
\newcommand{\dndy}         {\ensuremath{\mathrm{d}N_\mathrm{ch}/\mathrm{d}y}\xspace}
\newcommand{\dndeta}       {\ensuremath{\mathrm{d}N_\mathrm{ch}/\mathrm{d}\eta}\xspace}
\newcommand{\avdndeta}     {\ensuremath{\langle\dndeta\rangle}\xspace}
\newcommand{\dNdy}         {\ensuremath{\mathrm{d}N_\mathrm{ch}/\mathrm{d}y}\xspace}
\newcommand{\Npart}        {\ensuremath{N_\mathrm{part}}\xspace}
\newcommand{\Ncoll}        {\ensuremath{N_\mathrm{coll}}\xspace}
\newcommand{\dEdx}         {\ensuremath{\textrm{d}E/\textrm{d}x}\xspace}
\newcommand{\RpPb}         {\ensuremath{R_{\rm pPb}}\xspace}

\newcommand{\nineH}        {$\sqrt{s}~=~0.9$~Te\kern-.1emV\xspace}
\newcommand{\seven}        {$\sqrt{s}~=~7$~Te\kern-.1emV\xspace}
\newcommand{\twoH}         {$\sqrt{s}~=~0.2$~Te\kern-.1emV\xspace}
\newcommand{\twosevensix}  {$\sqrt{s}~=~2.76$~Te\kern-.1emV\xspace}
\newcommand{\five}         {$\sqrt{s}~=~5.02$~Te\kern-.1emV\xspace}
\newcommand{\twosevensixnn}{$\sqrt{s_{\mathrm{NN}}}~=~2.76$~Te\kern-.1emV\xspace}
\newcommand{\fivenn}       {$\sqrt{s_{\mathrm{NN}}}~=~5.02$~Te\kern-.1emV\xspace}
\newcommand{\LT}           {L{\'e}vy-Tsallis\xspace}
\newcommand{\GeVc}         {Ge\kern-.1emV/$c$\xspace}
\newcommand{\MeVc}         {Me\kern-.1emV/$c$\xspace}
\newcommand{\TeV}          {Te\kern-.1emV\xspace}
\newcommand{\GeV}          {Ge\kern-.1emV\xspace}
\newcommand{\MeV}          {Me\kern-.1emV\xspace}
\newcommand{\GeVmass}      {Ge\kern-.2emV/$c^2$\xspace}
\newcommand{\MeVmass}      {Me\kern-.2emV/$c^2$\xspace}
\newcommand{\lumi}         {\ensuremath{\mathcal{L}}\xspace}

\newcommand{\ITS}          {\rm{ITS}\xspace}
\newcommand{\TOF}          {\rm{TOF}\xspace}
\newcommand{\ZDC}          {\rm{ZDC}\xspace}
\newcommand{\ZDCs}         {\rm{ZDCs}\xspace}
\newcommand{\ZNA}          {\rm{ZNA}\xspace}
\newcommand{\ZNC}          {\rm{ZNC}\xspace}
\newcommand{\SPD}          {\rm{SPD}\xspace}
\newcommand{\SDD}          {\rm{SDD}\xspace}
\newcommand{\SSD}          {\rm{SSD}\xspace}
\newcommand{\TPC}          {\rm{TPC}\xspace}
\newcommand{\TRD}          {\rm{TRD}\xspace}
\newcommand{\VZERO}        {\rm{V0}\xspace}
\newcommand{\VZEROA}       {\rm{V0A}\xspace}
\newcommand{\VZEROC}       {\rm{V0C}\xspace}
\newcommand{\Vdecay} 	   {\ensuremath{V^{0}}\xspace}

\newcommand{\ee}           {\ensuremath{e^{+}e^{-}}} 
\newcommand{\pip}          {\ensuremath{\pi^{+}}\xspace}
\newcommand{\pim}          {\ensuremath{\pi^{-}}\xspace}
\newcommand{\kap}          {\ensuremath{\rm{K}^{+}}\xspace}
\newcommand{\kam}          {\ensuremath{\rm{K}^{-}}\xspace}
\newcommand{\pbar}         {\ensuremath{\rm\overline{p}}\xspace}
\newcommand{\kzero}        {\ensuremath{{\rm K}^{0}_{\rm{S}}}\xspace}
\newcommand{\lmb}          {\ensuremath{\Lambda}\xspace}
\newcommand{\almb}         {\ensuremath{\overline{\Lambda}}\xspace}
\newcommand{\Om}           {\ensuremath{\Omega^-}\xspace}
\newcommand{\Mo}           {\ensuremath{\overline{\Omega}^+}\xspace}
\newcommand{\X}            {\ensuremath{\Xi^-}\xspace}
\newcommand{\Ix}           {\ensuremath{\overline{\Xi}^+}\xspace}
\newcommand{\Xis}          {\ensuremath{\Xi^{\pm}}\xspace}
\newcommand{\Oms}          {\ensuremath{\Omega^{\pm}}\xspace}
\newcommand{\degree}       {\ensuremath{^{\rm o}}\xspace}

\begin{titlepage}
\PHyear{2022}       
\PHnumber{257}      
\PHdate{17 November}  

\title{Investigation of K$^{+}$K$^{-}$ interactions via femtoscopy in Pb--Pb collisions at {\ensuremath{\pmb{\sqrt{s_{\rm NN}}}}}{\bf ~=~2.76}~TeV at the LHC}
\ShortTitle{Investigation of $\rm K^+ K^-$ interactions via femtoscopy}   

\Collaboration{ALICE Collaboration\thanks{See Appendix~\ref{app:collab} for the list of Collaboration members}}
\ShortAuthor{ALICE Collaboration} 
\justify
\begin{abstract}
  Femtoscopic correlations of non-identical charged kaons ($\rm K^+ K^-$) are studied in Pb--Pb collisions at
  a center-of-mass energy per nucleon--nucleon collision
  $\sqrt{s_{\mathrm{NN}}} =2.76$ TeV by ALICE at the LHC. One-dimensional $\rm K^+ K^-$ correlation functions are analyzed in three centrality classes and
  eight intervals of particle-pair transverse momentum. 
The Lednick\'y and Luboshitz interaction model 
used in the $\rm K^+ K^-$ analysis includes
 the final-state Coulomb interactions
between kaons and the final-state interaction through $a_{0}$(980) and $f_{0}$(980) resonances.
The mass of $f_{0}$(980) and coupling were extracted from the fit to $\rm K^+ K^-$
correlation functions
using the femtoscopic technique.
The measured
mass and width of the $f_{0}$(980) resonance are consistent with other published measurements.
The height of the $\phi$(1020) meson peak
present
in the $\rm K^+ K^-$ correlation function rapidly decreases with increasing source radius, qualitatively in agreement with an inverse volume dependence.
A phenomenological fit to this trend suggests that the $\phi$(1020) meson yield is dominated by particles produced directly from the hadronization of the system. The small fraction subsequently produced by final-state interactions could not be precisely quantified with data presented in this paper and will be assessed in future work.
\end{abstract}
\end{titlepage}

\setcounter{page}{2} 

\justify
\section{Introduction}

Femtoscopy is a tool for measuring the space--time geometry of the particle emission region in
high\mbox{-}energy
collisions of protons and ions~\cite{LL82,lis05_b}.
It is based on the measurement of two-particle momentum correlation functions (CF) which
are determined by final-state interactions (FSI) between the emitted particles and effects of quantum statistics
in case of identical species~\cite{GGLP,LL82,lis05_a}. The technique was traditionally used to determine the size of the emission region
and its dependence on the particle-pair transverse
momentum and transverse mass, and on the event
multiplicity~\cite{NA49,ALICE_one_dimension}.
Recently, there has been a great interest in studying the interaction
of particles using femtoscopy methods
along with a parameterization of the size of the particle emitting source~\cite{ALICE:2018ysd,ALICE:2019gcn,ALICE:2021cpv,ALICE:2021cyj,ALICE:2021njx,ALICE:2019buq,ALICE:2019hdt,ALICE:2019eol,ALICE:2020mfd,Fabbietti:2020bfg}
to analyze the measured correlation functions.
There has also been interest in the femtoscopic correlations of pairs of non-identical kaons involving a neutral kaon $\Kzs\rm K^{\pm}$~\cite{ALICE_K0sKch_PbPb,ALICE_K0sKch_pp,ALICE:2021ovd}.
An important complement to these studies is the measurement
of the $\KpKm$ correlations.
The existing results in this area are rather scarce~\cite{NA49, Lidrych:2016uoh, STARKpKm}.
This is due to the complexity of the measurements and the subsequent complicated analysis.
In comparison to identical kaons, the interaction between K$^+$ and K$^-$ in the final state is much more complex.
It includes the Coulomb interaction and the strong interaction through
the near-threshold $f_0$(980) ($I$=0 isospin state) and $a_0$(980) ($I$=1
isospin
state) resonances,
and the strong p-wave FSI through the $\phi$(1020) meson.

The properties of the scalar mesons $a_0$(980) and  $f_0$(980), discovered in the mid-1960s and in the beginning of the 1970s, respectively~\cite{Rosenfeld:1965,Armenteros:1965zz,Protopopescu:1973sh,HYAMS1973134,Binnie:1973wkc}, are still subject of research.
The idea of the nature of these mesons as a quark--antiquark pair~\cite{Chen:2003za} is supplemented by a state in the form of a
K${\mathrm{\overline K}}$ 
molecule~\cite{Janssen:1994wn} and even a tetraquark state ~\cite{Jaffe.PhysRevD.15.267}.
Recently, in Ref.~\cite{ALICE:2021ovd},
it has been shown that the study of the magnitude of the correlation strength ($\lambda$)
in $\Kzs\Kzs$  and  $\Kzs\rm K^{\pm}$ pairs allows one to conclude that the observed difference in measured $\lambda$ is compatible with the $a_0$(980) resonance being a tetraquark state.
However, it should be noted that, according to the results of the latest ALICE work on $f_0$(980) in pp collisions at $\sqrt{s}=5.02$ TeV~\cite{ALICE:2022.f0.980}, the model
descriptions assuming a tetraquark (u$\overline{\rm u}$s$\overline{\rm s}$),
K${\mathrm{\overline K}}$ molecule, and s$\overline{\rm s}$
disagree with the experimental measurement.

The first measurement of $\KpKm$ correlations was carried out for Pb--Pb collisions
at the CERN SPS~\cite{NA49}.
It was shown that the theoretical $\KpKm$ correlation functions
calculated
by using a
finite-size Coulomb wave function with the radius extracted from identical kaon correlations were noticeably greater than the measured ones.
However, by taking into account the contribution due to
strong interactions, a reasonably good description of the data was obtained at small
relative momenta.

Preliminary results from the analysis of unlike-sign kaon femtoscopic correlations in Au--Au collisions at $\sqrt{s_{\mathrm{NN}}} =200$ GeV were reported by the STAR Collaboration~\cite{Lidrych:2016uoh, STARKpKm}.
The experimental one-dimensional
$\KpKm$ correlation function in terms of the invariant momentum difference
was compared with the theoretical prediction based on the Lednick\'y--Luboshitz approach~\cite{LL82}.
The measured $\KpKm$ CF could be described by the theoretical calculations using
a Gaussian function to model the source with the size parameters extracted
from the fit of the correlation function of identical charged kaons.
To account for additional physical effects not included in the theoretical function, the calculated correlation function ${\rm CF}^{\rm theor}$ was scaled according to ${\rm CF}=({\rm CF}^{\rm theor} - 1)\lambda -1$, where the correlation strength parameter
$\lambda$ was obtained from the fit to the like-sign kaon correlation function.
The STAR study showed that the model could qualitatively reproduce
the general structure of the measured
$\KpKm$ correlation function both at low relative momenta
$q<200$ MeV/$c$, where correlations are determined by the interplay of the Coulomb and s-wave strong interactions, and in the $\phi$(1020) resonance region.

In this work, $\KpKm$ femtoscopic correlations are studied for the first time in Pb--Pb collisions at a center-of-mass energy per nucleon--nucleon collision
$\sqrt{s_{\mathrm{NN}}} =2.76$ TeV at the CERN Large Hadron Collider (LHC) by the ALICE Collaboration~\cite{ALICE_LHC}. 
The physics goals of the present study are as follows:
1) extraction of the $f_0$(980) mass and coupling parameters based on the fit to the $\KpKm$ correlation function using
the Lednick\'y--Luboshitz model~\cite{LL82}
and on the assumption that the source radii of $\KpKm$ pairs are the same as those of
K$^{\pm}$K$^{\pm}$~\cite{ALICE_one_dimension}; 2) test of the $a_0$(980) mass and coupling parameters used in the $\Kzs$$\rm K^{\pm}$ femtoscopy study~\cite{ALICE_K0sKch_PbPb,ALICE_K0sKch_pp}; 3) investigation of how the height of the $\phi$(1020) peak in the CF changes with the source radius
in order to shed light on the nature of the production of the $\phi$(1020) meson in heavy-ion collisions.

The organization of this article
is as follows. In Sec.~\ref{sec_Data_selection},
the event and track selection criteria
are described. In Sec.~\ref{sec_Analysis_technique}, the theoretical and experimental details of the correlation functions and the fitting procedure are discussed. The results of the analysis are shown in Sec.~\ref{sec_Results_and_discussion}, and a summary is provided in Sec.~\ref{sec_Summary}.

\section{Event and charged-particle reconstruction and selection} \label{sec_Data_selection}

The analysis presented in this paper
used a sample of about 40 million minimum bias Pb--Pb collisions at $\sqrt{s_{\rm NN}}=2.76$~TeV collected with the ALICE detector 
in the LHC Run 1 period (2009--2013).
Monte Carlo (MC) simulations were used for correcting the obtained CFs for track momentum resolution.
In the simulations, 
particles from Pb--Pb collision events were generated with the HIJING~\cite{Wang:1991hta} general-purpose event generator and were propagated
through the ALICE detector using the GEANT3~\cite{Brun:1994aa} transport code.
The total number of MC events used in this analysis was about 4 millions.
Most of the event and track selection criteria in the current analysis are the same as in~\cite{ALICE_one_dimension}.

Events were classified according to their centrality determined using the measured signal amplitudes in the V0 detectors~\cite{V02011,ALICE:2013mez,ALICE:2013axi},
which consist of two arrays of scintillator counters
installed on each side of the interaction point and covering
the pseudorapidity intervals 
$2.8 \; \textless \; \eta \; \textless \; 5.1$
(V0A
) and \mbox{$-3.7 \; \textless \; \eta \; \textless \; -1.7$}
(V0C)~\cite{ALICE:2013axi}.
Charged particles were reconstructed and identified with the detectors located within a
solenoidal magnet that provides a uniform field of 0.5 T along the beam direction.
Charged particle tracking was performed using the Inner Tracking System (ITS)~\cite{ALICE:2010tia} and the Time Projection Chamber (TPC)~\cite{Alme:2010ke}.
The ITS consists of six cylindrical layers of silicon detectors, located at radii between 4 and 43 cm.
The ITS and TPC cover the pseudorapidity range $| \eta | < 0.9$ 
for all vertices located within
the interaction diamond~\cite{ALICE_LHC}.
The ITS provides high spatial resolution in determining the primary (collision) vertex and the distance of closest approach (DCA) of a track to the primary vertex. 
The primary-vertex position
along the beam direction ($z$ coordinate in the ALICE reference frame)
was required to be within $\pm$10~cm from the center
of the ALICE detector to ensure uniform tracking performance.
The TPC is the main part of the ALICE apparatus and was designed to
track and identify charged particles
in the high particle-density environment of heavy-ion collisions at the LHC.
The TPC is a 5 m long cylindrical gas detector with a volume close to 90 m$^3$ and with full acceptance in the pseudorapidity range $| \eta | < 0.9$.
The particle momenta were determined using tracks reconstructed with the TPC and constrained to originate from the primary vertex.
In order to reduce the number of secondaries, primary tracks were selected based on DCA to the primary vertex.
Additional track selections based on the quality of the track momentum fit and the number of detected space points in the TPC were used.
Each track was required to have at least 80 (out of a maximum of 159) associated space-points in the TPC. 
Track pairs sharing
more than 5\% of TPC clusters were rejected~\cite{ALICE_one_dimension}.
Particle identification (PID) was carried out using both the TPC and the Time of Flight (TOF)~\cite{Akindinov:2013tea} detectors in the pseudorapidity range $|\eta|<0.8$.
The TOF is a cylindrical detector with
a radius of about 3.7 m.
The total area of the active part of the TOF is about 141 m$^2$.
The main unit of the ALICE TOF detector is the Multigap Resistive Plate Chamber (MRPC) strip detector. The usable area of each MRPC is about
$120 \times 7.4$ cm$^2$. The ALICE TOF array was assembled from 1593 MRPC strips, subdivided into 18 azimuth sectors.
For TPC PID, a parametrized \mbox{Bethe--Bloch} formula for a particle with a given charge, mass, and momentum was used to calculate the expected specific 
energy loss ($ {\rm d}E/{\rm d}x $) in the detector.
The deviation between the measured and expected $ {\rm d}E/{\rm d}x $ values was required to be within a certain number of standard deviations ($N_{\sigma,\rm TPC}$) relative to the $ {\rm d}E/{\rm d}x $ resolution of the TPC~\cite{ALICE:2005vhb}. 
A similar $N_{\sigma,\rm TOF}$ selection was applied for particle identification with the TOF. In this case, the expected time of flight for a particle with a given mass was calculated from the track length and momentum measured with the tracking detectors. 
A detailed description of the particle identification is given in~\cite{ALICE:2014sbx}. 
The selection criteria which were used for kaon selection in the TPC and TOF are shown in Table~\ref{tab:KKcuts}.
\begin{table}
\centering
\caption{Charged kaon selection criteria.}
\begin{tabular}{ll}
\hline\hline
    $p_{\rm T}$  & $0.14<p_{\rm T}<1.5$ GeV/$c$ \\ 
    $|\eta|$ & $< 0.8$ \\ 
    $\rm DCA_{\rm transverse}$ to primary vertex & $< 2.4$ cm \\ 
    $\rm DCA_{\rm longitudinal}$ to primary vertex & $< 3.0$ cm \\ 
    $N_{\sigma,\rm TPC}$ (for $p < 0.4$~GeV/$c$) & $< 2$ \\
    $N_{\sigma,\rm TPC}$ (for $0.4 < p < 0.45$~GeV/$c$) & $< 1$ \\ 
    $N_{\sigma,\rm TPC}$ (for $p > 0.45$~GeV/$c$) & $< 3$ \\
    $N_{\sigma,\rm TOF}$ (for $0.45 < p < 0.8$ GeV/$c$) & $< 2$ \\
    $N_{\sigma,\rm TOF}$ (for $0.8 < p < 1.0$ GeV/$c$) & $< 1.5$ \\
    $N_{\sigma,\rm TOF}$ (for $1.0 < p < 1.5$ GeV/$c$) & $< 1.0$ \\
    Number of track points in TPC  & $\geq$80 \\
    $\chi^2/N_{\rm clusters}$ of the track fit   & $\leq$4 \\
\hline\hline
\end{tabular}
\label{tab:KKcuts}
\end{table}

\begin{figure}[h]
  \begin{center}
\includegraphics[width=0.48\textwidth]{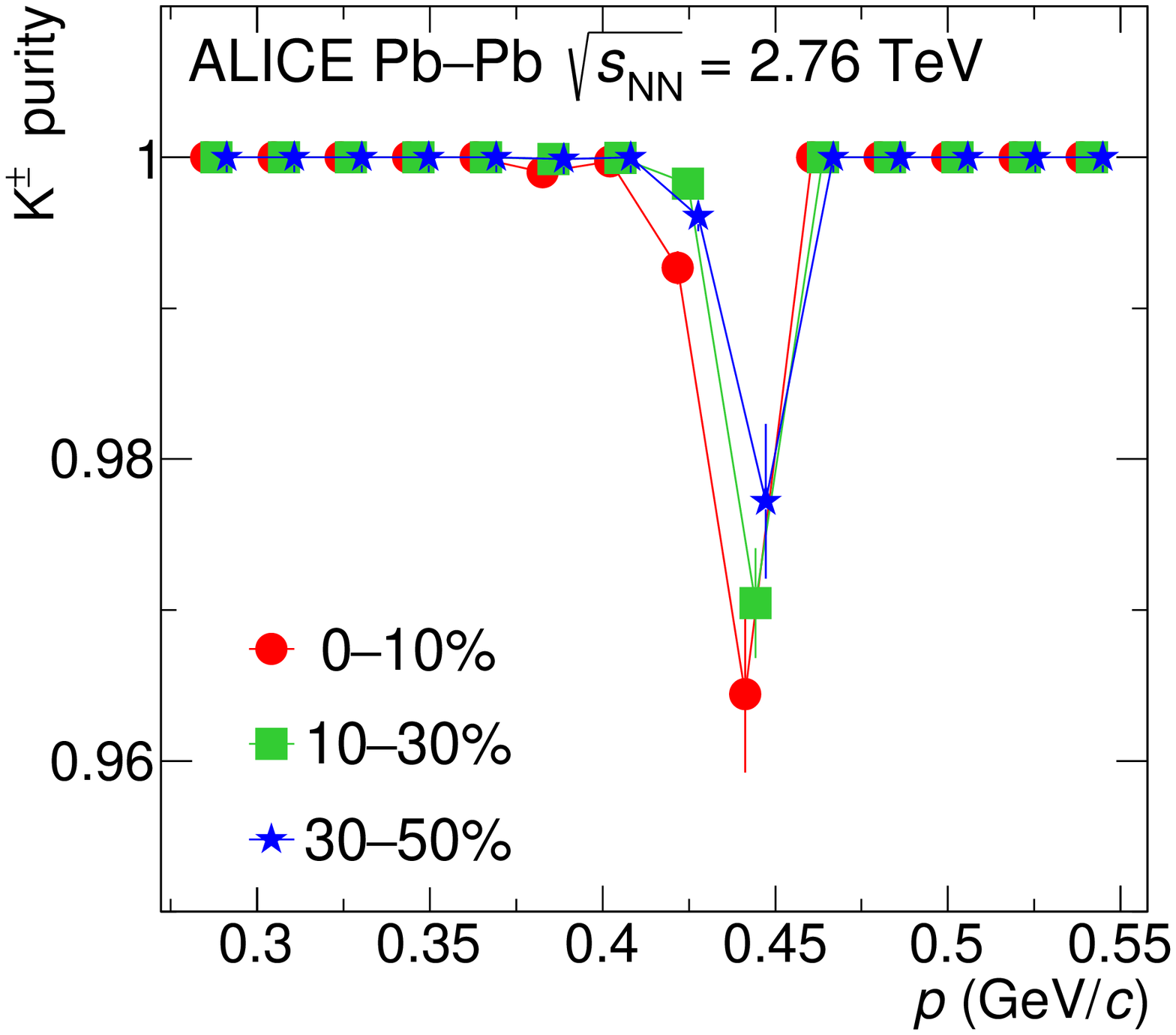}
\includegraphics[width=0.48\textwidth]{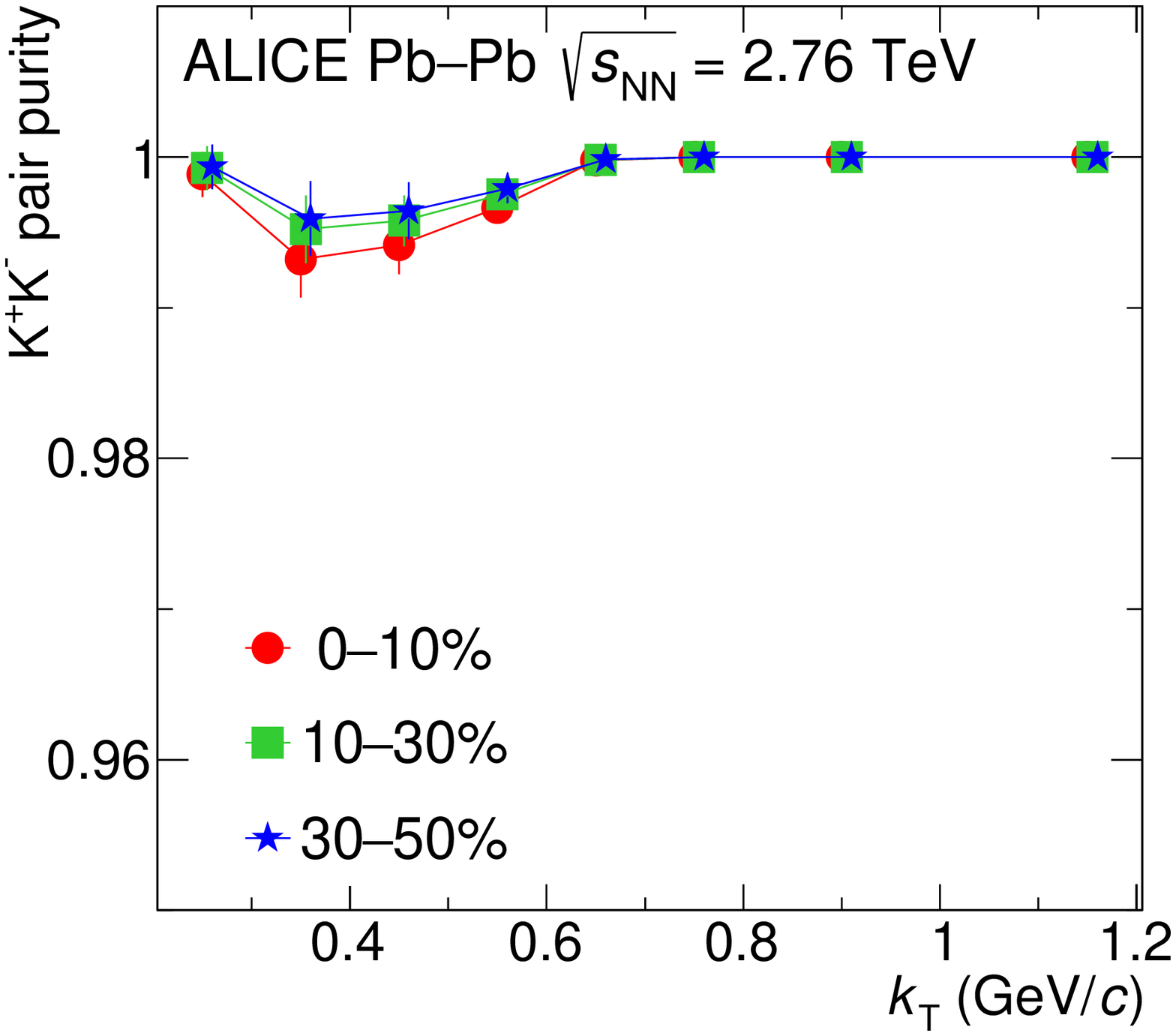}
  \end{center}
  \caption{
    Left:
    Single-kaon purity
    as a function of
    particle momentum
    $p$ for 0--10\%, 10--30\%, and 30--50\% centrality intervals.
  Points are shifted along the $x$ axis for clarity.
  Right: K$^+$K$^-$ pair purity
    as a function of
    pair transverse momentum $k_{\rm T}$. Systematic uncertainties are shown by
    bars.
Statistical uncertainties are smaller than the size of the markers.
}
\label{fig:purity}
\end{figure}

The purity of kaons is larger than 99\%
for tracks with momentum greater than 0.45 GeV/$c$~\cite{ALICE_one_dimension}.
To estimate the charged kaon purity for $p<0.45$\,GeV/$c$, the
measured $ {\rm d}E/{\rm d}x $ distribution was used~\cite{ALICE:2019kno}.
First,
the measured $ {\rm d}E/{\rm d}x $ distributions in track momentum intervals were considered, and the contributions of electrons, pions, kaons, and protons were parametrized via Gaussian fits.
Next, an estimate of the purity of kaons for momentum \mbox{ $p<0.45$\,GeV/$c$ } was made using this parametrization.
The estimated single kaon purity
as a function of momentum $p$ is shown in Fig.~\ref{fig:purity}~(left panel)
for different centrality intervals.
The obtained values of purity decrease
from semi-peripheral (30--50\%) to central (0--10\%) collisions.
The resulting kaon pair purity
as a function of pair transverse momentum $k_{\rm T}=|\vec{p}_{\rm T,1}+\vec{p}_{\rm T,2}|/2$ for different centralities is shown in Fig.~\ref{fig:purity}~
(right panel).
The pair purity distribution is wider and its values are larger on average than for the single-kaon purity. The value of the pair purity is higher than 99\% for $\KpKm$ pairs in the considered $k_{\rm T}$ interval.
The main contamination for $\KpKm$ pairs comes from $\gamma\rightarrow$e$^+$e$^-$ conversions.
It should be noted that to reduce this effect, the identification of kaons with the TOF starts when the charged kaon momentum is larger
than 0.45 GeV/$c$ instead of 0.5 GeV/$c$ as it was in the identical kaon femtoscopy analysis published in~\cite{ALICE_one_dimension}.
In addition,
a more stringent selection on $N_{\sigma}$ was applied in the momentum interval where the
contamination from
e$^+$e$^-$
pairs is expected to be large
(see~Table~\ref{tab:KKcuts}).

\section{Analysis technique} \label{sec_Analysis_technique}

Two-particle momentum correlations are defined as
$C(\vec{q})=A(\vec{q})/B(\vec{q})$, where $A(\vec{q})$
is the measured distribution of same-event pair momentum difference $\vec{q}=\vec{p}_1-\vec{p}_2$,
$\vec{p}_{1}$ and $\vec{p}_{2}$ are the momentum of
the first and second particle in the pair,
and $B(\vec{q})$ is the reference distribution of pairs from mixed events.
The mixed-event pair distribution was obtained by
mixing particles from events with similar centrality and vertex positions along the beam direction.
The correlation function is measured as a function of $q=\sqrt{|\vec{q}|^2 - q_0^2}$, where $q_0=E_1-E_2$ is determined by the
energies $E_1$, $E_2$ of the correlating particles. The correlation function is normalized to unity such that $C(q)\rightarrow1$ in the absence of a correlation signal.

\subsection{Theoretical description of K$^+$K$^-$ correlation function} \label{TheorCF}

This analysis studies femtoscopic correlations of particles produced in Pb--Pb collisions using the two-particle correlation function.
The measured K$^+$K$^-$ CF was fitted with a theoretical correlation function calculated within the Lednick\'y--Lyuboshitz approach~\cite{LL82,Led1998,LedBraz2007,LedReso}. The K$^{+}$ and K$^{-}$ particles were
assumed to be correlated in the final state due to the Coulomb interaction, to strong interactions through the near-threshold resonances $a_0$(980) and $f_0$(980), and to the p-wave strong interaction through the $\phi$(1020) meson resonance~\cite{LedReso}.

In the calculations, the correlations are conveniently expressed as a function of the single particle momentum in the pair rest frame (PRF), $k^*= |\vec{k}^*|$. Note that, in the case of pairs of particles with equal masses, $k^*$ is related to the momentum difference $q$ as $k^*=q/2$. For particle production occurring at a small enough phase-space density, the correlations of two particles emitted with small $k^*$ are dominated
by the effects of their mutual final-state interaction
and, if particles under consideration are identical, by quantum statistics.
These correlations depend on the PRF temporal ($t^*$) and spatial ($\vec{r}^*$) separation of the particle emission points.
Usually, one can neglect the temporal separation and in such equal-time approximation~\cite{LL82,Lednicky:2005tb} these effects are described
by properly symmetrized wave function $\Psi$. 
Assuming sufficiently smooth behavior of single-particle spectra in a narrow correlation region of small $k^*$ (smoothness assumption)~\cite{pod89_b},
one can write the K$^+$K$^-$ correlation function at a given $\vec{k}^*$
and the total pair three-momentum $P=|\vec{P}|$ as
\begin{equation}\label{eq:cfkoo}
C_{\rm FSI}(\vec{k}^*,\vec{P})= \int {\rm d}\vec{r}^*
\sum \limits_{\alpha'} S_{P}^{\alpha^\prime}(\vec{r}^*,\vec{k^*})\left| \Psi_{-\vec{
k}^*}^{\alpha^{\prime}\alpha}(\vec{r}^*)
\right|^2,
\end{equation}
where the sum is done
over the two intermediate channels $\alpha = $K$^+$K$^-$ and $\beta =$  K$^0{\mathrm{\overline K}^0}$, denoted by the index $\alpha^\prime$.
It is implied that particles are produced in a complex process with equilibrated spin and isospin projections. The separation distribution (source function)  $S_{P}^{\alpha^\prime}(\vec{r}^*,\vec{k^*})$ is then independent of these projections so that its channel index $\alpha^\prime$ can be omitted.
Assuming possible position--momentum correlations at particle freeze-out, the source function can be parametrized as~\cite{LedReso}
\begin{equation}\label{eq:fitgaus}
S_P(\vec{r^*},\vec{k^*}) = \exp \Bigg(-\frac{r^{*2}}{4R^2} + b\vec{r^*} \vec{k^*} \Bigg)/[8\pi^{3/2}R^3\exp(b^2{k^*}^{2}R^2)],
\end{equation}
where $R$ is the Gaussian source radius, and
$b$ is a $\vec{r^*}$~--~$\vec{k^*}$ correlation parameter.
Typically, $b\sim0.25$~\cite{LedReso}, the $\vec{r^*}$~--~$\vec{k^*}$ correlation in the low $k^*$ region
($k^* < 1/R$)
can be neglected,
and Eq.~(\ref{eq:fitgaus}) reduces to the usual spherically symmetric Gaussian parametrization.

Outside the range of the strong interaction potential and at a sufficiently small $k^*$, one may account only for the s-wave strong interaction
and write~\cite{Lednicky:2005tb,LedBraz2007}
\begin{eqnarray}
&&\Psi^{\alpha\alpha}_{-\vec{k}^*}(\vec{r}^*) = {\cal N}(\eta) \left[ {\rm e}^{-i\vec{k}^*\vec{r}^*} F(-i\eta,1,i\xi) + f_c^{\alpha\alpha}(k^*) \frac{\widetilde{G}(\rho,\eta)}{r^*}\right],\\
&&\Psi^{\beta\alpha}_{-\vec{k}^*}(\vec{r}^*) = {\cal N}(\eta) f_c^{\beta\alpha}(k^*) \sqrt{\frac{\mu_{\beta}}{\mu_\alpha}}
\frac{{\rm e}^{ik^*_\beta r^*}}{r^*},
\end{eqnarray}
where ${\cal N}(\eta)={\rm e}^{i\delta_c(\eta)} \sqrt{A_c(\eta)}$, $\eta=(k^*a)^{-1}$,
$a$ is the two-particle Bohr radius including the sign of the interaction (for K$^+$K$^-$ $a=$-109.6\,fm),
$\rho=k^*r^*$, $\mu_\alpha=m_{\rm K^+}/2$, $k^*=k^*_\alpha$ and $\mu_{\beta}=m_{\rm K^0}/2$,
$k^*_\beta$ are the respective reduced masses and K$^+$ and K$^0$ momenta in PRF,
$\delta_c=\arg\Gamma(1+i\eta)$ is the Coulomb s-wave phase shift,
$A_c(\eta) = 2 \pi \eta [{\rm e}^{2\pi\eta}-1]^{-1}$ is the Coulomb penetration (Gamow) factor,
$F$ is the confluent hypergeometric function, and $\widetilde{G}$ is a combination of the regular and singular s-wave Coulomb functions~\cite{Gmitro:1986ay}.
The s-wave scattering amplitudes $f_c^{\alpha^\prime \alpha}$
due to the short-range interaction renormalized by the long-range Coulomb forces are the elements of a 2$\times$2 matrix~\cite{Lednicky:2005tb,LedBraz2007,STAR_K0K0}
\begin{equation} \label{eq:fhatc}
\hat{f}_c = \left( \hat{K}^{-1} - i\hat{k}_c\right)^{-1}.
\end{equation}
Here $\hat{K}$ is a symmetric matrix and $\hat{k}_c$ is a diagonal matrix in the channel representation $k_c^{\alpha\alpha}=A_c(\eta)k^*-2ih(\eta)/a$, $k^{\beta\beta}_c=k_2^*$ with $k_2^*$ being the kaon momentum in PRF of the inelastic channel, where the function $h(\eta)$ is expressed through the digamma function $\psi$ as $h(\eta)=[\psi(i\eta)-\psi(-i\eta)+\ln\eta^2]/2$.

The elements of the $\hat{K}^{-1}$ matrix in the channel flavor
representation $\alpha = $K$^+$K$^-$, $\beta =$K$^0{\mathrm{\overline K}^0}$ are expressed through elements $K^{-1}_{\rm I}$ of the diagonal matrix $\hat{K}^{-1}$ in the representation of the channel isospin $I = 0, 1$ as~\cite{LedBraz2007}
\begin{eqnarray} \label{eq:Khat}
&&(\hat{K}^{-1})^{\alpha\alpha} = (\hat{K}^{-1})^{\beta\beta} = \frac{1}{2}(K_0^{-1}+K_1^{-1}),\\
&&(\hat{K}^{-1})^{\beta\alpha} = (\hat{K}^{-1})^{\alpha\beta} = -\frac{1}{2}(K_0^{-1}-K_1^{-1}).\label{eq:Khat1}
\end{eqnarray}

The latter are assumed to be dominated by the isoscalar $f_0$(980) and isovector $a_0$(980) resonances, so
\begin{equation} \label{eq:fampl2}
K_0(k^*)=\frac{\gamma_{f_0 \rightarrow  {\rm K\overline{K}}}}
{m^2_{f_0} - s - i \gamma_{f_0 \rightarrow \pi \pi}k_{\pi \pi} },
\end{equation}
\begin{equation} \label{eq:fampl3}
K_1(k^*)=\frac{\gamma_{a_0 \rightarrow  {\rm K\overline{K}}}}
{m^2_{a_0} - s - i\gamma_{a_0 \rightarrow \pi \eta}k_{\pi \eta} },
\end{equation}
where $s=4(m^2_{\rm K}+k^{*2})$, $m_{a_0}$ and $m_{f_0}$ are the masses of the $a_0$(980) and $f_0$(980) resonances, respectively (see Table~\ref{tab:coupar}), $\gamma_{f_0 \rightarrow {\rm K\overline{K}}} $, $\gamma_{f_0 \rightarrow \pi \pi} $ and $\gamma_{a_0 \rightarrow  {\rm K\overline{K}}} $, $\gamma_{a_0 \rightarrow \pi \eta} $ are the respective couplings, and $k_{\pi \eta}$, $k_{\pi \pi}$ are the decay pion momenta in the respective channels.

To take into account the deviation of the spherical waves from the true scattered waves in the inner region of the short-range potential, a correction $\Delta C_{{\rm K}\overline{\rm K}}$ should be applied (see Eq.~(153) in~\cite{Lednicky:2005tb})
\begin{equation}\label{eq:Dc}
\Delta C_{{\rm K}\overline{\rm K}} = -2\pi S_P(0,k^*) A_c(\eta) [|f_c^{\alpha\alpha}|^2 d_0^{\alpha\alpha} + |f_c^{\beta\alpha}|^2 d_0^{\beta\beta}+ 2 \mathfrak{R}(f_c^{\alpha\alpha} f_c^{\beta\alpha*}) d_0^{\beta\alpha}],
\end{equation}
where $d_0^{\alpha^\prime\alpha}= 2 \mathfrak{R} {\rm d}(\hat{K}^{-1})^{\alpha^\prime \alpha}/{\rm d}{k^*}^2$.

An additional contribution to the K$^+$K$^-$ correlation function due to the p-wave strong interaction through the $\phi$(1020) meson resonance was also taken into consideration.
The usual femtoscopic correlation formalism of 
Eq.~(\ref{eq:cfkoo}) using the smoothness assumption at small $k^*$ should be modified at the $\phi \to $~K$^+$K$^-$ decay momentum $k_0 = 127$ MeV/$c$
to account for substantial $\vec{r}-\vec{k}$ correlations quantified by the
parameter $b\sim0.25$ in Eq.~(\ref{eq:fitgaus}). As a result, the $\phi$(1020) contribution to the correlation function is exponentially suppressed by the normalization factor in Eq.~(\ref{eq:fitgaus}) and, neglecting a small Coulomb correction, becomes~\cite{LedReso}
\begin{equation}\label{eq:Dcphi}
C_\phi^{\rm FSI} = S_P(0,k^*)\frac{6\pi}{k^*} \left[\mathfrak{R} \left[\frac{{\rm d}f_\phi^{\alpha\alpha}}{{\rm d}k^*}\right]
+ \sum_{\alpha^\prime} 2 \mathfrak{I}\left[k^*_{\alpha^\prime} f_\phi^{\alpha^\prime\alpha*} \frac{{\rm d}f_\phi^{\alpha^\prime \alpha}}{{\rm d}k^*}\right]\right],
\end{equation}
where $f_\phi^{\alpha^\prime\alpha} =
\pm
   [(\Gamma_{\alpha^\prime}/k^*_{\alpha^\prime})(\Gamma_\alpha/k^*)]^{1/2} m_\phi/(m_\phi^2-s-im_\phi\Gamma)$, $\Gamma= \sum_{\alpha^\prime} \Gamma_{\alpha^\prime} + \Gamma^\prime$ is the total $\phi$(1020) width,
   $\Gamma^\prime= 0.168 \Gamma$
is the partial width of the $\phi$(1020) decays to the channels other than the K$\overline{\rm K}$ ones, and 
$\Gamma_{\alpha^\prime} \sim k^{*3}_{\alpha^\prime}$.
The sign $\pm$ corresponds to $\alpha'= \alpha$ and $\beta$, respectively.
The expression for $f_{\phi}^{\alpha^\prime\alpha}$ follows from Eqs.~(\ref{eq:fhatc})-(\ref{eq:Khat1}) with the substitution
$\hat{K}^{-1} \rightarrow \hat{k}^L\hat{K}^{-1}\hat{k}^L$ in
Eqs. (\ref{eq:Khat}) and (\ref{eq:Khat1})
due to non-zero orbital angular momentum $L=1$. Here $\hat{k}$ is a diagonal matrix in the channel flavor representation, $k^{\alpha\alpha}=k^*$,  $k^{\beta\beta}=k_2^*$;
in the channel isospin representation,
$k^{00}=k^{11}=(k^{\alpha\alpha}+k^{\beta\beta})/2$,
$k^{01}=k^{10}=(k^{\alpha\alpha}-k^{\beta\beta})/2$.
The matrix $\hat{k}\hat{K}^{-1}\hat{k}$ is diagonal in the channel isospin representation
$(\hat{k}^{-1}\hat{K}\hat{k}^{-1})^{00}= \gamma_{\phi \to \rm K\overline{K}}/[m_\phi^2 -s -i m_\phi \Gamma']$
and $(\hat{k}^{-1}\hat{K}\hat{k}^{-1})^{11} \rightarrow 0$.
Equation~(\ref{eq:Dcphi}) has the same structure as the s-wave correction
in the inner
region~\cite{Led1998} and can thus be substituted by Eq.~(\ref{eq:Dc}) multiplied by the p-wave factor $2L+1=3$.
By neglecting the difference between the K$^+$K$^-$ and K$^0\overline{\rm K^0}$ channel momenta, the $\phi$(1020) contribution can be rewritten as
\begin{equation}\label{eq:cphifsi}
 C_\phi^{\rm FSI} \doteq \frac{12\pi S_P(0,k^*)|f^{\alpha \alpha}|^2} {(\mu_\alpha \Gamma_\alpha/k^*)}
     \doteq \frac{6\pi S_P(0,k^*)(\Gamma-\Gamma^\prime)}{\mu_\alpha k_0|(k_0^2-k^{*2})/\mu_\alpha-i\Gamma|^2}.
\end{equation}
Here the usage of the non-relativistic
Breit--Wigner expression in the last equality
is motivated by the narrow $\phi$(1020) width allowing one to neglect the momentum dependence of $\Gamma_\alpha$.

The total correlation function is determined
as a sum of the s-wave term described in Eqs.~(\ref{eq:cfkoo})-(\ref{eq:Dc}) and
the p-wave term FSI described by Eq.~(\ref{eq:Dcphi}) and the direct $\phi$(1020) meson production
\begin{equation} \label{eq:Cfsi}
  \begin{split}
  C(p_1,p_2) =  1 + \lambda ( C_{a_0,f_0}^{\rm FSI}(p_1,p_2) +  C_{\phi}(p_1,p_2) ), \\
  C_{\phi}(p_1,p_2) = a_{\rm direct} C_{\phi}^{\rm direct}(p_1,p_2) +a_{\rm FSI} C_{\phi}^{\rm FSI}(p_1,p_2),
  \end{split}
\end{equation}
where $C_{\phi}^{\rm direct}$ can be described by a non-relativistic Breit--Wigner function~\cite{PhiALICEPaper}, $C_{\phi}^{\rm FSI}$ is calculated from Eq.~(\ref{eq:Dcphi}), and $a_{\rm direct}$ and $a_{\rm FSI}$ are coefficients that determine the ratio of $\phi$(1020) mesons produced directly and due to FSI, respectively.
Within the experimental accuracy, $C_{\phi}^{\rm FSI}$ can be described by a non-relativistic Breit--Wigner function.
Therefore, this function
was used to fit $C_{\phi}$ (where $C_{\phi}$ is the non-relativistic Breit--Wigner function) instead of the two separate terms $a_{\rm direct} C_{\phi}^{\rm direct}$ and $a_{\rm FSI} C_{\phi}^{\rm FSI}$. The ratio of the contributions of $\phi$(1020) meson production from the direct mechanism $a_{\rm direct}$ and from the coalescence mechanism $a_{\rm FSI}$ is discussed in Sec.~\ref{phi_radius}.
\begin{table}[!t]
  \caption{$a_0$(980) and $f_0$(980) square masses (in GeV$^2$/$c^4$) and coupling parameters (in GeV).}
\begin{center}
  \begin{tabular}{lcccccc}
\hline\hline
 Model & $m^{2}_{f_0}$ & $m^2_{a_0}$ & $\gamma^{{\phantom{1}}^{\phantom{1}}}_{f_0 \rightarrow  {\rm K^+K^-}} $ & $\gamma_{f_0 \rightarrow \pi \pi} $& $\gamma_{a_0 \rightarrow  {\rm K^+K^-}} $& $\gamma_{a_0 \rightarrow \pi \eta} $  \\
\hline
Martin~\cite{Martin} &0.9565&0.9487&0.792 &0.199 &0.333 &0.222 \\
Antonelli~\cite{Antonelli} &0.9467 &0.9698 &2.763 &0.5283 &0.4038 &0.3711 \\
Achasov1~\cite{Achasov1} &0.9920 &0.9841 &1.305 &0.2684 &0.5555 &0.4401 \\
Achasov2 ~\cite{Achasov2}&0.9920 &1.0060 &1.305 &0.2684 &0.8365 &0.4580 \\
\hline\hline
\end{tabular}
\end{center}
\label{tab:coupar}
\end{table}

\subsection{The K$^+$K$^-$ correlation function and fitting procedure} \label{ExpCF}%

\begin{figure}[t]
\begin{center}
  \includegraphics[width=0.99\textwidth]{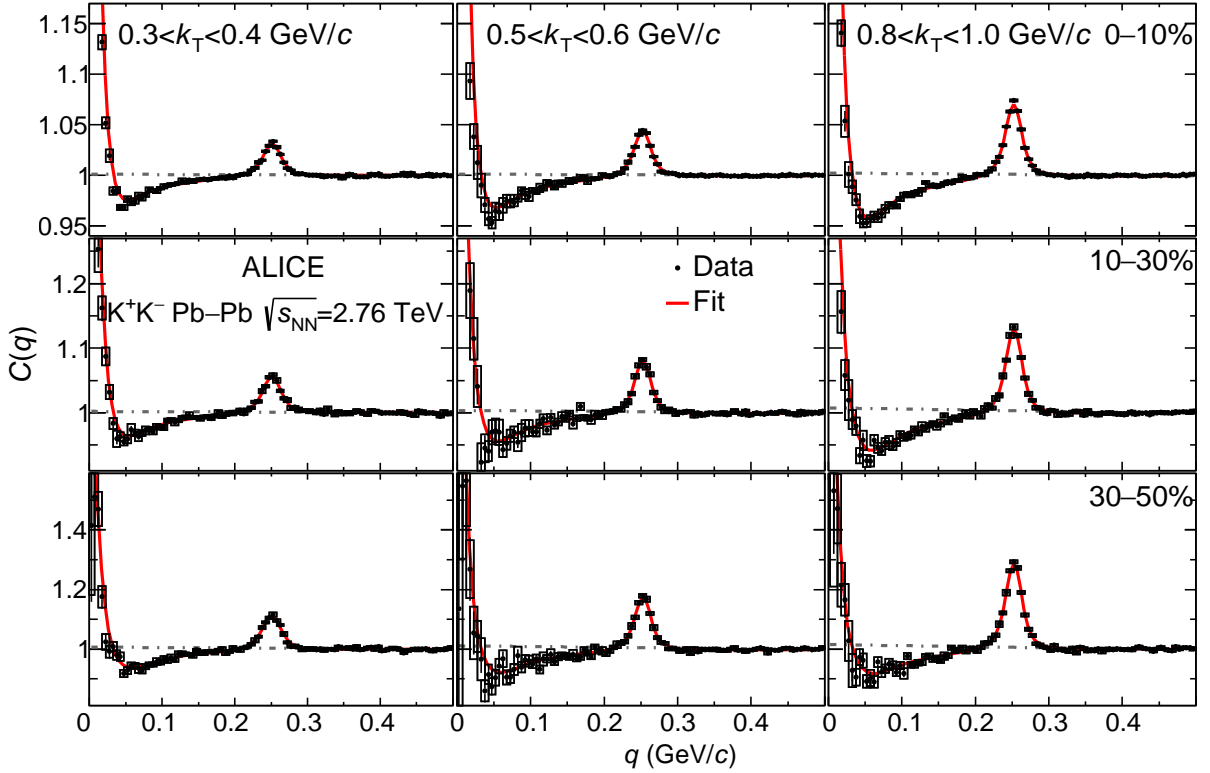} \caption{
The K$^+$K$^-$ experimental correlation functions corrected for non-flat baselines according to Eq.~(\ref{eq:baseline})
as a function of pair relative momentum $q$. The CFs are presented in three
centrality classes (rows): 0--10\%, 10\mbox{--}30\%, and 30--50\% and three pair transverse momentum $k_{\rm T}$ bins (columns): (0.3--0.4), (0.5--0.6) and (0.8\mbox{--}1.0)~GeV/$c$.
Statistical (bars) and systematic (boxes) uncertainties are shown. 
The red line shows the
fit of the CF with the Lednick\'y--Lyuboshitz parametrization (Eq.~(\ref{eq:Cfsi})) using free parameters (mass and couplings) for $f_0$(980) and Achasov~\cite{Achasov2} parameters for $a_0$(980) in the $0<q <  0.5$ GeV/$c$ range.
The dashed-dotted lines correspond to the baseline from Eq.~(\ref{eq:baseline}).
}
\label{fig:CF3x3}
\end{center}
\end{figure}

The correlation functions were measured in eight pair transverse momentum $k_{\rm T}$ intervals:
(0.2--0.3), (0.3--0.4), (0.4--0.5), (0.5--0.6), (0.6--0.7), (0.7--0.8), (0.8--1.0), and (1.0--1.3) GeV/$c$ and three centrality classes: 0--10\%, 10--30\%, 30--50\%.
As an example, the K$^+$K$^-$ correlation functions for three of these eight $k_{\rm T}$ bins in different centrality classes are shown in Fig.~\ref{fig:CF3x3}.
The measured correlation functions were normalized to unity in the region of $0.35 < q < 0.5$ GeV/$c$
and corrected for momentum resolution as described in Sec.~\ref{sect:MR}.
In this figure, one can see the main features of the K$^+$K$^-$ femtoscopic correlation function: Coulomb attraction at very small $q$ ($q<0.05$ GeV/$c$)
resulting in $C(q) > 1$,
suppression ($C(q) < 1$) due to strong final-state interactions via the formation
of the near-threshold resonances $a_0$(980) and $f_0$(980) in $0.05<q<0.2$ GeV/$c$, and the narrow $\phi$(1020) resonance peak at $q$ around 0.25 GeV/$c$.

The measured correlation function was corrected for the non-flat baseline $D$ before the fit. The baseline is fitted in a wide $q$ range ($0.35 < q < 1.0$ GeV/$c$) using a first-order polynomial function
\begin{equation}\label{eq:baseline}
D(q) = \kappa (1+aq),
\end{equation}
where $\kappa$ is a normalization factor, and $a$ is a free parameter of the fit.
The observed non-flat baseline effect is almost negligible for the most central collisions and at low $k_{\rm T}$, while it becomes significant at low multiplicities and high transverse momenta.
The non-flat baseline could be associated with the manifestation of mini-jets in peripheral collisions.
A similar effect was observed for K$^{\pm}$K$^{\pm}$ in Pb--Pb collisions at $\sqrt{s_{\rm NN}}=2.76$ TeV~\cite{ALICE_one_dimension}.
The changing trend of the slope
in different $k_{\rm T}$ intervals could be reproduced qualitatively by
HIJING simulations but the magnitude of the slope in the simulation was
different than the one in the data in all considered $k_{\rm T}$ intervals.
Therefore, the MC was not used to estimate the non-flat baseline effect in this analysis.

The fit of the calculated correlation function to
the measured distributions allows one to constrain the masses and coupling parameters
of the $a_0$(980) and $f_0$(980) resonances.
The $a_0$(980) parameters were fixed using the Achasov model~\cite{Achasov2} in the K$^0_{\rm S}$K$^{\pm}$ femtoscopic correlation
analysis of Pb--Pb collisions at $\sqrt{s_{\rm NN}}=2.76$ TeV and of pp collisions at $\sqrt{s}=7$ TeV~\cite{{ALICE_K0sKch_PbPb},{ALICE_K0sKch_pp}}.
Therefore, only the parameters of the $f_0$(980) resonance were studied in this work.
Three possible sets of values of $f_0$(980) parameters proposed by theoretical models (Marin~\cite{Martin}, Antonelli~\cite{Antonelli}, and Achasov~\cite{{Achasov1},{Achasov2}}, see Table~\ref{tab:coupar})
were considered.
The source radii for K$^+$K$^-$ pairs obtained by using all these theoretical models are inconsistent
with the source radii parameters obtained in the
identical sign K$^{\pm}$K$^{\pm}$ analysis~\cite{ALICE_one_dimension}, while there are no physical reasons
for this difference.
Thus, in this study the parameters of the $f_0$(980) resonance
have been estimated by treating them as free parameters in the fits and constraining
the source radii of K$^+$K$^-$ pairs to be consistent with those obtained from the analysis of identical sign K$^{\pm}$K$^{\pm}$ correlations.
The parameter $\lambda$ for K$^+$K$^-$ does not necessarily have to be equal to the parameter $\lambda$ for K$^{\pm}$K$^{\pm}$. In the case of identical charged kaon correlations, the $\lambda$ parameter decreases with increasing $k_{\rm T}$, which can be attributed to
a non-Gaussian shape of the source~\cite{ALICE_one_dimension}.
Instead, the K$^+$K$^-$ correlation function is determined by the contribution of the Coulomb and strong FSI, which are not very sensitive to
a possible non-Gaussian shape of the source.

The fits of the K$^+$K$^-$ experimental correlation function for three different $k_{\rm T}$ and centrality intervals with the Lednick\'y--Lyuboshitz parametrization using free parameters (mass and couplings) for $f_0$(980) and Achasov~\cite{Achasov2} parameters for $a_0$(980) are shown in Fig.~\ref{fig:CF3x3}.
Systematic uncertainties of CF were estimated using correlation functions obtained with different magnetic field orientations in the detector. As seen from Fig.~\ref{fig:CF3x3},
the $f_0$(980) FSI parametrization obtained in this analysis gives an excellent description
of the data in the interval $0 < q < 0.35$ GeV/$c$, where its contribution is relevant.
The corresponding $\chi^2$/ndf are in the range from 1 to 2.

\subsection{Finite momentum resolution}\label{sect:MR}
Finite track momentum resolution causes the reconstructed relative momentum $q_{\rm rec}$ of a pair to differ from the true value $q_{\rm true}$.
This is accounted for through the use of the response matrix $M(q_{\rm true},q_{\rm rec})$ generated with HIJING simulations
of Pb--Pb collisions at $\sqrt{s_{\rm NN}}$ = 2.76 TeV.
To account for this effect, the theoretical CF can be smeared through the response matrix according to~\cite{ALICE:KaonLambdaBuxton}
\begin{equation}\label{eq:mqtrueqrec}
C(q_{\rm rec}) = \frac{ \sum\limits_{q_{\rm true}} C(q_{\rm true}) M(q_{\rm true},q_{\rm rec})}{\sum\limits_{q_{\rm true}} M(q_{\rm true},q_{\rm rec})}.
\end{equation}
This smearing was applied directly in the fit of the measured CF to the theoretical one.
The momentum resolution effect was included in the theoretical CF used to fit the data.
In general, the momentum resolution reduces the height of the correlation function peak and makes it wider.

\subsection{Systematic uncertainties}\label{Systematic_uncertainties}
Possible sources and estimated values of systematic uncertainties of
the source radius $R$ and correlation strength $\lambda$ parameters 
are presented in this section.
Systematic uncertainties of the mass and coupling parameter of the $f_0$(980) meson are also discussed.
The total systematic uncertainty
\begin{equation}
\Delta_{\rm sys} = \sqrt{\sum_{i} (\Delta_{\rm sys}^{i})^2}
\label{eq:SysErrSum}
\end{equation}
was taken as the square-root of the quadratic sum of all systematic contributions $\Delta_{\rm sys}^{i} = |y_{0} - y_{\rm var}^i|$ from the fit and the selection criteria (Table~\ref{tab:systerrKch}).
Here $y_{0}$ refers to the value obtained with the default criteria 
(central value), $y_{\rm var}^i$ is a value with some variation $i$ of particle selections and fit criteria.
The Barlow factor~\cite{Barlow} was also considered 
to estimate a statistical significance level of each deviation as
\begin{equation}
B = \frac{ | y_{0} - y^{i}_{\rm var} | }{ \sqrt{\sigma_{0}^2 + \sigma_{\rm var}^2 - 2\rho\sigma_{0}\sigma_{\rm var}} },
\label{eq:Barlow}
\end{equation}
where $\sigma_{0}$ indicates the statistical uncertainty of the central 
value, $\sigma_{\rm var}$ is the statistical uncertainty of $y^{i}_{\rm var}$, $\rho$ characterizes correlation between $y_{0}$ and $y^{i}_{\rm var}$.
The variation $i$ is included in the systematic uncertainty evaluation if $B$ is larger than unity.

\begin{table}[h]
\centering
\caption{Summary of relative systematic uncertainties on $R$ and $\lambda$ parameters. The symbol '-' means that the contribution from the given source is negligible. The ranges reported for each specific source and for the total uncertainty reflect the fact that the uncertainty values depend on centrality and $k_{\rm T}$ intervals. Only systematic uncertainties whose statistical significance level exceeds 68\% according to the Barlow criterion were considered.}
 \begin{tabular}{lll}
 \hline\hline
 Sources of systematic uncertainty& $R$ (\%) & $\lambda$ (\%)  \\ \hline
 {Single particle selection} & 0--10  & 0--12 \\
 {Purity}                    & --     &  0--0.5 \\
{Baseline fit range}                 & {0--4} & 0--3 \\
 {Momentum resolution}       & 3--9   & 4--25 \\
{Total (quad. sum)}         & 3--14  & 4--28 \\
 \hline\hline
 \end{tabular}
\label{tab:systerrKch}
\end{table}

The effect of the track selections
was investigated by varying the criteria shown in Table~\ref{tab:KKcuts}.
The DCA and PID selection values were varied by $\pm$10\%.
These variations resulted up to a 10--12\%
contribution to the systematic uncertainties for the $R$ and $\lambda$ parameters (see Table~\ref{tab:systerrKch}). The residual contamination from other particle species in the K$^+$K$^-$ pair signal was found to have a minimal effect on the extracted parameters because of the high purity (see Fig.~\ref{fig:purity}) of selected kaons, which was better than 99\% for a pair of kaons.

The systematic effect due to the choice of the baseline fit range was estimated by varying the $q$ interval in which the fit is performed. The standard fit range of the baseline was $0.35<q< 1.0$ GeV/$c$, and the upper limit of $q$ was changed to 0.7 and 1.3 GeV/$c$.
The estimated uncertainty depends on the centrality and the $k_{\rm T}$ interval, and its maximum and minimum values are reported in Table~\ref{tab:systerrKch}.

Changing the range of the fit slightly changes the femtoscopic parameters extracted from the correlation function.
However, after correcting the correlation function for the non-flat baseline, the influence of the fit range variation was found to be negligible.

The effect of finite momentum resolution on femtoscopic radii and $\lambda$ parameters and the related systematic uncertainty were studied by modifying the width of momentum resolution distribution in the response matrix 
described in Sec.~\ref{sect:MR}.
The width of the $q_{\rm true}$ vs $q_{\rm rec}$ distribution determined by $M(q_{\rm true},q_{\rm rec})$ was varied maximally ($\pm$10\%) to account for its influence on the CF without distorting its shape.

The resulting systematic uncertainties are reported in Table~\ref{tab:systerrKch}.
The systematic uncertainties of the $\gamma_{f_0 \rightarrow  {\rm K\overline{K}}}$
and $\gamma_{f_0 \rightarrow \pi \pi}$ couplings are determined by their possible
maximal deviation from the default values providing the best CF fit under condition
of having close radii for $\rm K^+ K^-$ to those obtained for pairs of identical kaons,
and a constraint on
the  $\gamma_{f_0 \rightarrow  {\rm K\overline{K}}} / \gamma_{f_0 \rightarrow \pi \pi}$
ratio to be in accordance with the predictions of the
models~\cite{{Martin},{Antonelli},{Achasov1},{Achasov2}}.

\section{Results and discussion} \label{sec_Results_and_discussion}
\subsection{Source size and correlation strength}
\begin{figure}[h]
\begin{center}
\includegraphics[width=0.49\textwidth]{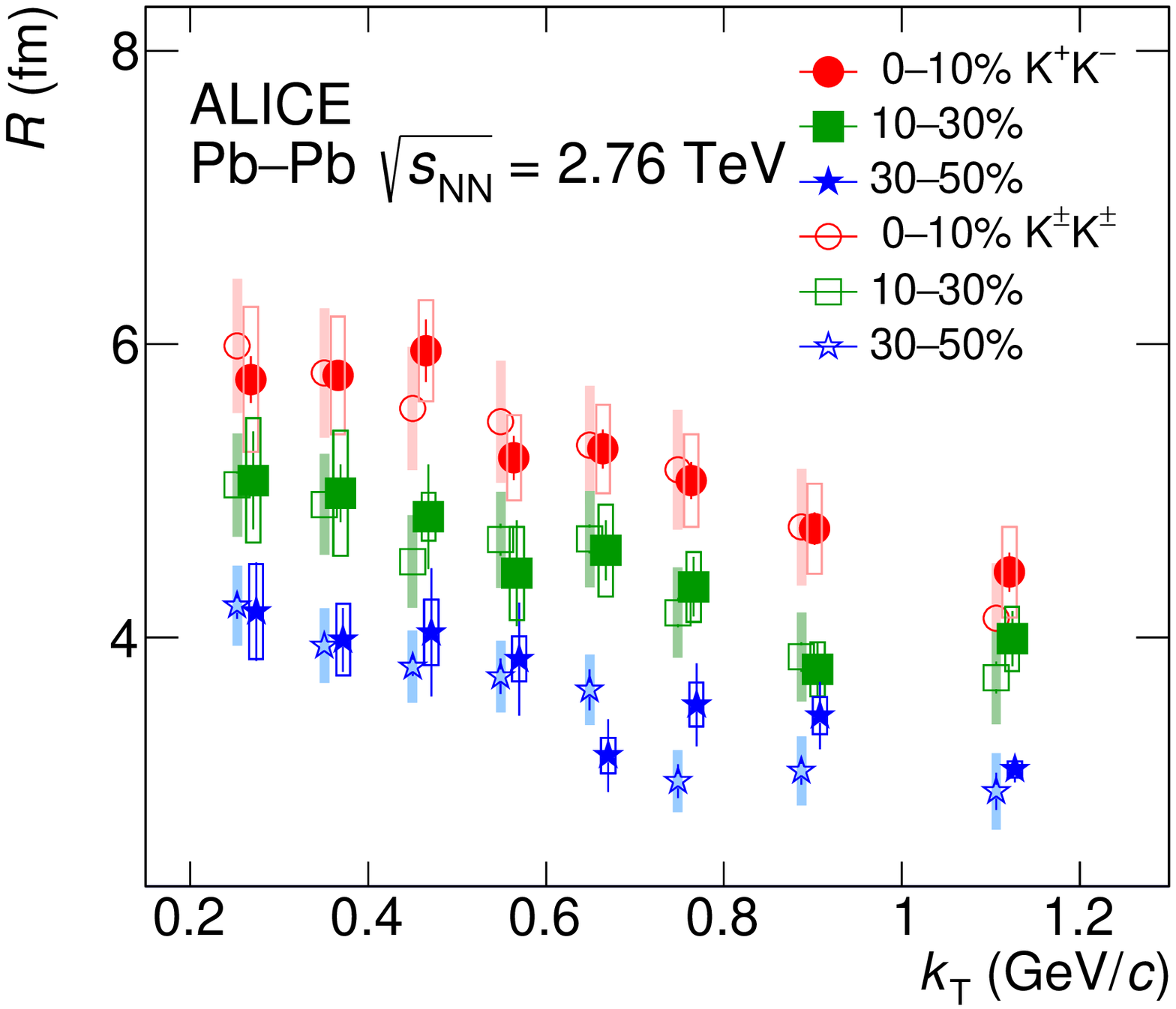}
\includegraphics[width=0.49\textwidth]{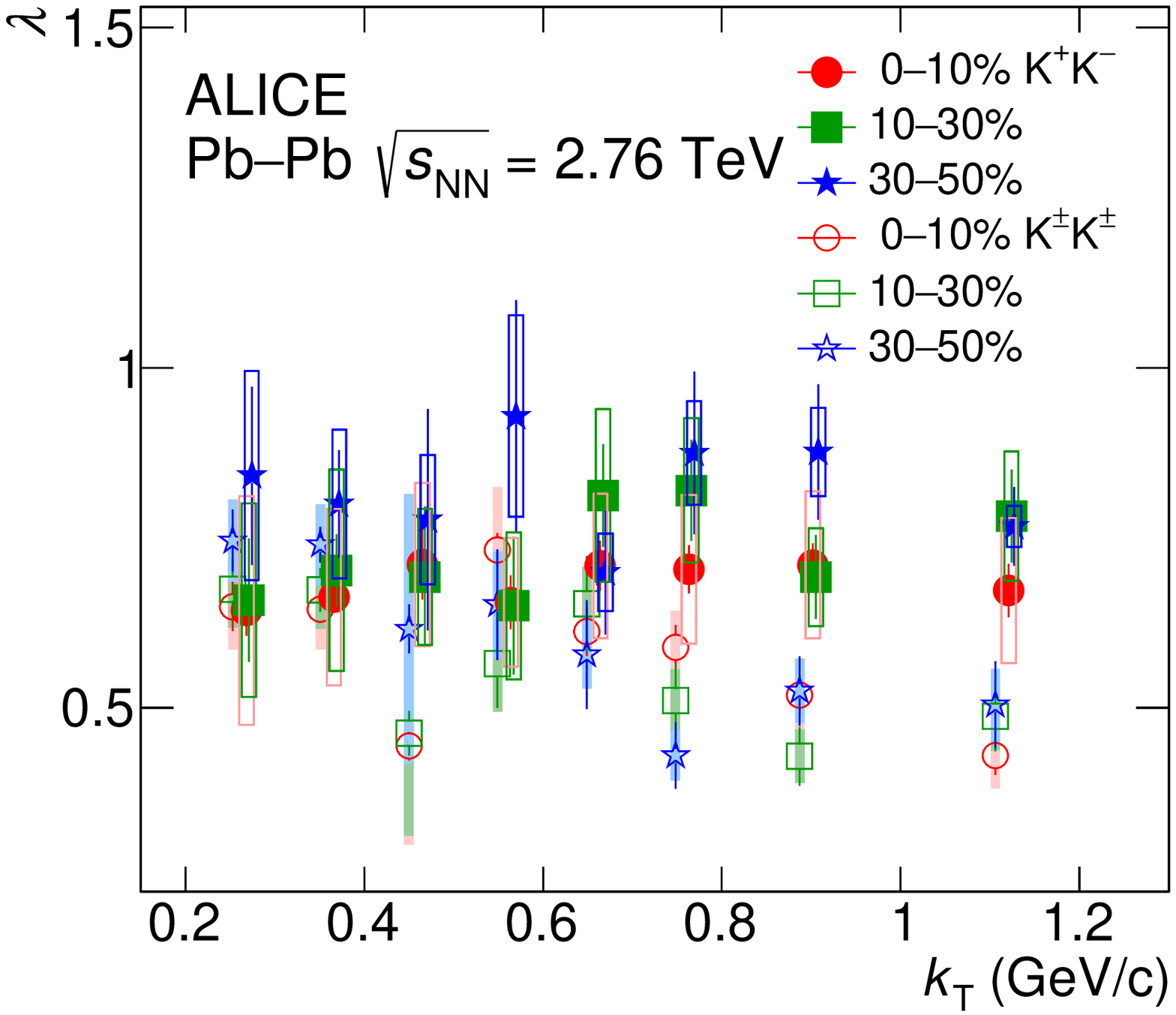}
\caption{
$R$ (left panel) and $\lambda$ (right panel) parameters as a function of pair transverse momentum $k_{\rm T}$ extracted in K$^+$K$^-$ analysis with free parameters (mass and couplings) for $f_0$(980) and Achasov~\cite{Achasov2} parameters for $a_0$(980).
The parameters are compared to those obtained for identical charged kaons~\cite{ALICE_one_dimension}.
Statistical (bars) and systematic (boxes) uncertainties are shown.}
\label{fig:Note_RLX_Fit_f0_Model50_a0-Achasov2}
\end{center}
\end{figure}

In the CF fit, the $R$ parameters for K$^+$K$^-$ correlations were constrained to be compatible with the corresponding parameters for
the same sign K$^{\pm}$K$^{\pm}$ pairs within statistical uncertainties since there are no physical reasons for them to be different as was discussed above.
Figure~\ref{fig:Note_RLX_Fit_f0_Model50_a0-Achasov2} shows the obtained radii $R$~(left panel) and correlation strengths $\lambda$~(right panel)
as a function of $k_{\rm T}$ for the 0--10\%, 10--30\%, and 30--50\% centrality intervals.
The parameters extracted from K$^+$K$^-$ correlations are compared to those obtained for K$^{\pm}$K$^{\pm}$ pairs.

The $R$ parameters from K$^+$K$^-$ correlations shown in Fig.~\ref{fig:Note_RLX_Fit_f0_Model50_a0-Achasov2} (left panel) are by construction consistent with those from identical kaon pairs, due to the constraint applied in the fit. This constraint was implemented by minimizing $\chi^2/N$ that was calculated for each $R$ value according to
\begin{equation}
\chi^2_R/N = \sum_{i=1}^{N} \frac {[R_i({\rm K}^{\pm}{\rm K}^{\pm}) -R_i({\rm K}^{+}{\rm K}^{-})]^2 }{\sigma^2_i},
\label{eq:chi2R}
\end{equation}
where $i$ in $R_i$ runs over eight $k_{\rm T}$ values for each centrality bin ($N$ = 8) and $\sigma_i = \sqrt{\sigma^2_{{\rm K}^{\pm}{\rm K}^{\pm}}+\sigma^2_{{\rm K}^{+}{\rm K}^{-}}}$ is the statistical uncertainty of the difference between the extracted radius parameters for K$^+$K$^-$ and K$^{\pm}$K$^{\pm}$ correlations.
The obtained $\chi^2_R/N$ values are 1.5 for 0--10\%, 0.5 for 10--30\% and 1.1 for 30--50\% centralities.

The correlation strength parameters for K$^+$K$^-$
pairs tend to be slightly
larger than those for K$^{\pm}$K$^{\pm}$ for $k_{\rm T}>$0.7~GeV/$c$. In particular, for K$^+$K$^-$ pairs the $\lambda$ parameter does not show the decreasing trend with increasing $k_{\rm T}$ that was found for identical charge K$^{\pm}$K$^{\pm}$ correlations.
This difference could be due to the fact that the decreasing trend in the identical sign result is due to a non-Gaussian shape of the source and to the fact that the  K$^+$K$^-$ pairs are less sensitive to it.
The values of $\lambda$ are about 0.7, i.e.
lower than the ideal value of unity. This can be due to the contribution of kaons from K$^{\ast}$ decays and from other long-lived resonances distorting the spatial kaon source distribution
with respect to an ideal Gaussian, which is assumed in the fit function~\cite{ALICE_one_dimension}.

Mass and coupling parameters for the $f_0$(980) meson were extracted in this analysis using Eq.~(\ref{eq:fampl2}) with the constraint on the K$^+$K$^-$ radii to be close to the corresponding K$^{\pm}$K$^{\pm}$ radii as was explained above.
The resulting mass and coupling parameters with statistical and systematic uncertainties are
$m_{f_0}$~= 967~$\pm$~3~$\pm$~7~${\rm MeV}/c^2$,
$\gamma_{f_0 \rightarrow  {\rm K\overline{K}}}$~= 0.34~$\pm$~0.07~$\pm$~0.10~${\rm GeV}$ ,
$\gamma_{f_0 \rightarrow \pi \pi}$~=~0.089~$\pm$~0.018~$\pm$~0.026~${\rm GeV}$.
The obtained $f_0$(980) mass is consistent within uncertainties with its PDG value ($m_{f_0}$~=~990~$\pm$~20~${\rm MeV}/c^2$)~\cite{PDG2020}.
The ratio of kaon to pion couplings is equal to
$\gamma_{f_0 \rightarrow  {\rm K\overline{K}}} / \gamma_{f_0 \rightarrow \pi \pi}$~=~3.82~$\pm$~1.07
and is consistent with those shown in Table~\ref{tab:coupar}, which are in the range from 4 to 5 for all models.
The full width of the $f_0$(980) meson estimated in this work is $\Gamma_{f_0}$ = 43.81~$\pm$~8.76~$\pm$~6.90~MeV/$c^2$ and is also consistent with the PDG value ($\Gamma_{f_0}$ = 10--100 MeV/$c^2$).

\subsection{The $\phi$(1020) meson peak height versus radius}\label{phi_radius}

\begin{figure}[t]
  \begin{center}
    \includegraphics[width=.49\linewidth]{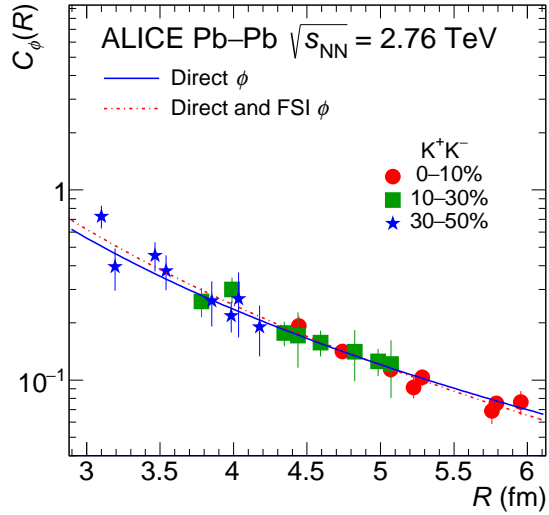}
  \end{center}
  \caption{Height of $\phi$(1020) meson peak ($C_\phi$) as a function of source radius $R$, for three centrality classes.
Statistical uncertainties are shown by bars. Systematic uncertainties are smaller than the size of the markers. Blue solid line corresponds to the fit of CF with
$C_\phi^{\rm direct} = {\rm const}/R^3$. Red dashed line corresponds to the fit with the second line in Eq.~(\ref{eq:Cfsi}).
}
  \label{fig:DeltaCFvsRinv}
\end{figure}

Figure~\ref{fig:CF3x3} shows that the height of the peak related to $\phi$(1020) meson decays depends on both $k_{\rm T}$ and centrality.
Since it is also known that the source radius changes with $k_{\rm T}$ and centrality,
it is interesting to investigate how the height of the peak changes with the radius.
The height of the $\phi$(1020) meson peak can be measured from the correlation function minus unity at a relative momentum $q$ that corresponds to the $\phi$(1020) meson mass as
\begin{equation} \label{eq:DeltaC}
  C_\phi = C(q=\sqrt{m_{\phi}^2-4m_{\rm K}^2}=2 k_0)-1.
\end{equation}
The $C_\phi$ value as a function of
the K$^+$K$^-$ radius is presented in Fig.~\ref{fig:DeltaCFvsRinv}.
It should be noted that it was corrected for both the $\lambda$ parameter magnitude and the momentum resolution.

It can be assumed that the mechanism of the $\phi$(1020) meson production consists of at least two processes.
The first is the direct production of $\phi$(1020) mesons
at the time of hadronization of the system.
The second is the regeneration via the K$^+$K$^-$ FSI
in the subsequent hadronic phase leading to resonance formation~\cite{LedReso}.

To estimate the contribution from directly produced $\phi$(1020) mesons
($a_{\rm direct}$
in Eq.~(\ref{eq:Cfsi})), the following reasoning can be applied.
The correlation function is defined as a ratio of the signal to the background.
According to statistical models of hadron production~\cite{Cleymans:2005xv}, the $\phi$(1020) meson yield (signal) is proportional to the volume of the $\phi$(1020) meson production region.
The combinatorial background is proportional to the square of the multiplicity, which in its turn is proportional to the source volume.
Therefore, the height of the $\phi$(1020) meson peak in the K$^+$K$^-$ correlation function is expected to rapidly decrease with increasing $R$ in qualitative accordance with an inverse volume dependence, i.e. as $1/R^3$.
Based on these arguments, it is possible to fit
the height of the $\phi$(1020) meson peak in the K$^+$K$^-$ correlation function
with $C_\phi^{\rm direct} = {\rm const}/R^3$ as shown in Fig.~\ref{fig:DeltaCFvsRinv}.
This fit function describes the data very well, indicating
that the FSI contribution
($a_{\rm FSI}$ in Eq.~(\ref{eq:Cfsi})) to the production of $\phi$(1020) mesons is quite small.

The influence of FSI can be estimated from simple model considerations.
The interaction of kaons in the final state depends on the separation of their
production points, which is affected by collective flow and resonance decays leading to the $\vec{r}$--$\vec{k}$ correlation characterized by the parameter $b$~\cite{LedReso}.
The ${\rm K^+K^-}$ FSI contribution could be approximately described by Eq.~(\ref{eq:Dcphi}) and Eq.~(4) in~\cite{LedReso}
\begin{equation}
  C_\phi^{\rm FSI} = \frac{{\rm const\times exp}(-b^2k_0^2R^2)}{R^3}.
\label{eq:rkFSI}
\end{equation}
Assuming that the direct production of $\phi$(1020) mesons is described by the inverse volume dependence $1/R^3$ while the FSI production is described by Eq.~(\ref{eq:rkFSI}), the K$^+$K$^-$ correlation function can be fit with the second line in Eq.~(\ref{eq:Cfsi}).
The parameter $b$ is fixed to the value about 0.25 obtained from the
Blast-Wave model estimation~\cite{LedReso}.
The results of the calculation are presented in Fig.~\ref{fig:DeltaCFvsRinv}.
The height of the $\phi$(1020) meson peak in the K$^+$K$^-$ correlation function is fit with
$C_\phi^{\rm direct} = {\rm const}/R^3$ (blue curve in Fig.~\ref{fig:DeltaCFvsRinv}). The constant is equal to 15.08$\pm$0.44.
The value of $\chi^2$/ndf=13.45/23=0.58. The results of the
fit of the data to the second line in Eq.~(\ref{eq:Cfsi})
are also shown in Fig.~\ref{fig:DeltaCFvsRinv}. The red curve corresponds to the fit with parameters of the fit
$a_{\rm direct}=$0.75$\pm$0.16, $a_{\rm FSI}=$0.25$\pm$0.16,
and $\chi^2/$ndf$=$11.07$/$21$=$0.53.
The resulting fraction of directly produced $\phi$(1020) mesons
$a_{\rm direct} C_\phi^{\rm direct}/ C_\phi$
varies from 0.7 to 0.8 with increasing $R$ and is within the range expected from the integrated hydrokinetic model~\cite{SinyukovNucPhys2017}.
Consequently, the possible fraction of $\phi(1020)$ mesons produced in FSI decreases from 0.3 to 0.2 with increasing $R$.

Figure~\ref{fig:DeltaCFvsRinv} shows that the resulting fits with Eq.~(\ref{eq:Cfsi}) and with simple $\sim1/R^3$ are very close to each other.
As a result of the present study, it can be concluded that the fraction of  $\phi$(1020) meson produced in FSI is difficult to distinguish from the directly produced $\phi$(1020) mesons, at least within the statistical precision of the data sample considered in this analysis.

\section{Summary} \label{sec_Summary}

In this article, the results of femtoscopic studies of non-identical charged kaon correlations in Pb--Pb collisions at \mbox{$\sqrt{s_{\mathrm {NN}}} = 2.76$~TeV} measured with the ALICE detector at the LHC are presented.
The K$^+$K$^-$ femtoscopic radii
were constrained to the corresponding parameters extracted from the analysis of identical charged kaon correlations in Pb--Pb collisions at the same collision energy.
The $a_0$(980) resonance mass and coupling parameters used in this work were fixed based on the $\Kzs$$\rm K^{\pm}$ femtoscopy study.
The parameters of the $f_0$(980) meson proposed by the Martin, Antonelli, and Achasov models did not provide a good description of the K$^+$K$^-$ correlation parameters
if the K$^+$K$^-$ source radii were required to be close to the corresponding K$^{\pm}$K$^{\pm}$ ones.
Therefore, the K$^+$K$^-$ correlation function was fitted using the $f_0$(980) mass and couplings as free parameters.
The extracted $f_0$(980) width 43.81$\pm$8.76$\pm$6.90~MeV/$c^2$ and mass 967$\pm$3$\pm$7~${\rm MeV}/c^2$ were found to be
consistent with the existing PDG world-average values.
The obtained values of the $f_0$(980) meson coupling parameters are $\gamma_{f_0 \rightarrow  {\rm K\overline{K}}} =$0.34$\pm$0.068$\pm$0.101~${\rm GeV}$ and
$\gamma_{f_0 \rightarrow \pi \pi} =$0.089$\pm$0.0178$\pm$0.026~${\rm GeV}$.
For the first time the parameters of a resonance ($f_0$(980) here) were obtained from femtoscopic measurements.
It was also shown that the height of the $\phi$(1020) meson peak in the $\rm K^+ K^-$ correlation function rapidly decreases with increasing $R$ as $1/R^3$ where $R$
is the radius of the particle emitting source.
A phenomenological fit to this trend suggests that the $\phi$(1020) meson yield is dominated by particles produced directly from the hadronization of the system.
The small fraction subsequently produced by FSI could not be precisely quantified with the data sample investigated in this analysis and will be assessed in future work.
It is difficult to estimate the relative contributions of $\phi$(1020) mesons from FSI and those produced directly in collisions
within the statistical precision of the existing data sample.
The
fraction of $\phi(1020)$ mesons produced in FSI could be estimated using the experimental data sample collected with the ALICE detector in the LHC Run 2 period, which is larger than the Run 1 sample used in the presented work, and is a subject of future research.


\newenvironment{acknowledgement}{\relax}{\relax}
\begin{acknowledgement}
\section*{Acknowledgements}

The ALICE Collaboration would like to thank all its engineers and technicians for their invaluable contributions to the construction of the experiment and the CERN accelerator teams for the outstanding performance of the LHC complex.
The ALICE Collaboration gratefully acknowledges the resources and support provided by all Grid centres and the Worldwide LHC Computing Grid (WLCG) collaboration.
The ALICE Collaboration acknowledges the following funding agencies for their support in building and running the ALICE detector:
A. I. Alikhanyan National Science Laboratory (Yerevan Physics Institute) Foundation (ANSL), State Committee of Science and World Federation of Scientists (WFS), Armenia;
Austrian Academy of Sciences, Austrian Science Fund (FWF): [M 2467-N36] and Nationalstiftung f\"{u}r Forschung, Technologie und Entwicklung, Austria;
Ministry of Communications and High Technologies, National Nuclear Research Center, Azerbaijan;
Conselho Nacional de Desenvolvimento Cient\'{\i}fico e Tecnol\'{o}gico (CNPq), Financiadora de Estudos e Projetos (Finep), Funda\c{c}\~{a}o de Amparo \`{a} Pesquisa do Estado de S\~{a}o Paulo (FAPESP) and Universidade Federal do Rio Grande do Sul (UFRGS), Brazil;
Bulgarian Ministry of Education and Science, within the National Roadmap for Research Infrastructures 2020¿2027 (object CERN), Bulgaria;
Ministry of Education of China (MOEC) , Ministry of Science \& Technology of China (MSTC) and National Natural Science Foundation of China (NSFC), China;
Ministry of Science and Education and Croatian Science Foundation, Croatia;
Centro de Aplicaciones Tecnol\'{o}gicas y Desarrollo Nuclear (CEADEN), Cubaenerg\'{\i}a, Cuba;
Ministry of Education, Youth and Sports of the Czech Republic, Czech Republic;
The Danish Council for Independent Research | Natural Sciences, the VILLUM FONDEN and Danish National Research Foundation (DNRF), Denmark;
Helsinki Institute of Physics (HIP), Finland;
Commissariat \`{a} l'Energie Atomique (CEA) and Institut National de Physique Nucl\'{e}aire et de Physique des Particules (IN2P3) and Centre National de la Recherche Scientifique (CNRS), France;
Bundesministerium f\"{u}r Bildung und Forschung (BMBF) and GSI Helmholtzzentrum f\"{u}r Schwerionenforschung GmbH, Germany;
General Secretariat for Research and Technology, Ministry of Education, Research and Religions, Greece;
National Research, Development and Innovation Office, Hungary;
Department of Atomic Energy Government of India (DAE), Department of Science and Technology, Government of India (DST), University Grants Commission, Government of India (UGC) and Council of Scientific and Industrial Research (CSIR), India;
National Research and Innovation Agency - BRIN, Indonesia;
Istituto Nazionale di Fisica Nucleare (INFN), Italy;
Japanese Ministry of Education, Culture, Sports, Science and Technology (MEXT) and Japan Society for the Promotion of Science (JSPS) KAKENHI, Japan;
Consejo Nacional de Ciencia (CONACYT) y Tecnolog\'{i}a, through Fondo de Cooperaci\'{o}n Internacional en Ciencia y Tecnolog\'{i}a (FONCICYT) and Direcci\'{o}n General de Asuntos del Personal Academico (DGAPA), Mexico;
Nederlandse Organisatie voor Wetenschappelijk Onderzoek (NWO), Netherlands;
The Research Council of Norway, Norway;
Commission on Science and Technology for Sustainable Development in the South (COMSATS), Pakistan;
Pontificia Universidad Cat\'{o}lica del Per\'{u}, Peru;
Ministry of Education and Science, National Science Centre and WUT ID-UB, Poland;
Korea Institute of Science and Technology Information and National Research Foundation of Korea (NRF), Republic of Korea;
Ministry of Education and Scientific Research, Institute of Atomic Physics, Ministry of Research and Innovation and Institute of Atomic Physics and University Politehnica of Bucharest, Romania;
Ministry of Education, Science, Research and Sport of the Slovak Republic, Slovakia;
National Research Foundation of South Africa, South Africa;
Swedish Research Council (VR) and Knut \& Alice Wallenberg Foundation (KAW), Sweden;
European Organization for Nuclear Research, Switzerland;
Suranaree University of Technology (SUT), National Science and Technology Development Agency (NSTDA), Thailand Science Research and Innovation (TSRI) and National Science, Research and Innovation Fund (NSRF), Thailand;
Turkish Energy, Nuclear and Mineral Research Agency (TENMAK), Turkey;
National Academy of  Sciences of Ukraine, Ukraine;
Science and Technology Facilities Council (STFC), United Kingdom;
National Science Foundation of the United States of America (NSF) and United States Department of Energy, Office of Nuclear Physics (DOE NP), United States of America.
In addition, individual groups or members have received support from:
Marie Sk\l{}odowska Curie, European Research Council, Strong 2020 - Horizon 2020 (grant nos. 950692, 824093, 896850), European Union;
Academy of Finland (Center of Excellence in Quark Matter) (grant nos. 346327, 346328), Finland;
Programa de Apoyos para la Superaci\'{o}n del Personal Acad\'{e}mico, UNAM, Mexico.

\end{acknowledgement}

\bibliographystyle{utphys}   
\bibliography{bibliography}

\newpage
\appendix

%
%

\section{The ALICE Collaboration}
\label{app:collab}
\begin{flushleft} 
\small

S.~Acharya\,\orcidlink{0000-0002-9213-5329}\,$^{\rm 125}$, 
D.~Adamov\'{a}\,\orcidlink{0000-0002-0504-7428}\,$^{\rm 86}$, 
A.~Adler$^{\rm 69}$, 
G.~Aglieri Rinella\,\orcidlink{0000-0002-9611-3696}\,$^{\rm 32}$, 
M.~Agnello\,\orcidlink{0000-0002-0760-5075}\,$^{\rm 29}$, 
N.~Agrawal\,\orcidlink{0000-0003-0348-9836}\,$^{\rm 50}$, 
Z.~Ahammed\,\orcidlink{0000-0001-5241-7412}\,$^{\rm 132}$, 
S.~Ahmad\,\orcidlink{0000-0003-0497-5705}\,$^{\rm 15}$, 
S.U.~Ahn\,\orcidlink{0000-0001-8847-489X}\,$^{\rm 70}$, 
I.~Ahuja\,\orcidlink{0000-0002-4417-1392}\,$^{\rm 37}$
% ...

\section*{Affiliation Notes}

$^{\rm I}$ Deceased\\
$^{\rm II}$ Also at: Max-Planck-Institut f\"{u}r Physik, Munich, Germany\\
$^{\rm III}$ Also at: Italian National Agency for New Technologies, Energy and Sustainable Economic Development (ENEA), Bologna, Italy\\
$^{\rm IV}$ Also at: Dipartimento DET del Politecnico di Torino, Turin, Italy\\
$^{\rm V}$ Also at: Department of Applied Physics, Aligarh Muslim University, Aligarh, India\\
$^{\rm VI}$ Also at: Institute of Theoretical Physics, University of Wroclaw, Poland\\
$^{\rm VII}$ Also at: An institution covered by a cooperation agreement with CERN\\

\section*{Collaboration Institutes}

$^{1}$ A.I. Alikhanyan National Science Laboratory (Yerevan Physics Institute) Foundation, Yerevan, Armenia\\
$^{2}$ AGH University of Science and Technology, Cracow, Poland\\
$^{3}$ Bogolyubov Institute for Theoretical Physics, National Academy of Sciences of Ukraine, Kiev, Ukraine\\
$^{4}$ Bose Institute, Department of Physics  and Centre for Astroparticle Physics and Space Science (CAPSS), Kolkata, India\\
$^{5}$ California Polytechnic State University, San Luis Obispo, California, United States\\
$^{6}$ Central China Normal University, Wuhan, China\\
$^{7}$ Centro de Aplicaciones Tecnol\'{o}gicas y Desarrollo Nuclear (CEADEN), Havana, Cuba\\
$^{8}$ Centro de Investigaci\'{o}n y de Estudios Avanzados (CINVESTAV), Mexico City and M\'{e}rida, Mexico\\
$^{9}$ Chicago State University, Chicago, Illinois, United States\\
$^{10}$ China Institute of Atomic Energy, Beijing, China\\
$^{11}$ Chungbuk National University, Cheongju, Republic of Korea\\
$^{12}$ Comenius University Bratislava, Faculty of Mathematics, Physics and Informatics, Bratislava, Slovak Republic\\
$^{13}$ COMSATS University Islamabad, Islamabad, Pakistan\\
$^{14}$ Creighton University, Omaha, Nebraska, United States\\
$^{15}$ Department of Physics, Aligarh Muslim University, Aligarh, India\\
$^{16}$ Department of Physics, Pusan National University, Pusan, Republic of Korea\\
$^{17}$ Department of Physics, Sejong University, Seoul, Republic of Korea\\
$^{18}$ Department of Physics, University of California, Berkeley, California, United States\\
$^{19}$ Department of Physics, University of Oslo, Oslo, Norway\\
$^{20}$ Department of Physics and Technology, University of Bergen, Bergen, Norway\\
$^{21}$ Dipartimento di Fisica, Universit\`{a} di Pavia, Pavia, Italy\\
$^{22}$ Dipartimento di Fisica dell'Universit\`{a} and Sezione INFN, Cagliari, Italy\\
$^{23}$ Dipartimento di Fisica dell'Universit\`{a} and Sezione INFN, Trieste, Italy\\
$^{24}$ Dipartimento di Fisica dell'Universit\`{a} and Sezione INFN, Turin, Italy\\
$^{25}$ Dipartimento di Fisica e Astronomia dell'Universit\`{a} and Sezione INFN, Bologna, Italy\\
$^{26}$ Dipartimento di Fisica e Astronomia dell'Universit\`{a} and Sezione INFN, Catania, Italy\\
$^{27}$ Dipartimento di Fisica e Astronomia dell'Universit\`{a} and Sezione INFN, Padova, Italy\\
$^{28}$ Dipartimento di Fisica `E.R.~Caianiello' dell'Universit\`{a} and Gruppo Collegato INFN, Salerno, Italy\\
$^{29}$ Dipartimento DISAT del Politecnico and Sezione INFN, Turin, Italy\\
$^{30}$ Dipartimento di Scienze MIFT, Universit\`{a} di Messina, Messina, Italy\\
$^{31}$ Dipartimento Interateneo di Fisica `M.~Merlin' and Sezione INFN, Bari, Italy\\
$^{32}$ European Organization for Nuclear Research (CERN), Geneva, Switzerland\\
$^{33}$ Faculty of Electrical Engineering, Mechanical Engineering and Naval Architecture, University of Split, Split, Croatia\\
$^{34}$ Faculty of Engineering and Science, Western Norway University of Applied Sciences, Bergen, Norway\\
$^{35}$ Faculty of Nuclear Sciences and Physical Engineering, Czech Technical University in Prague, Prague, Czech Republic\\
$^{36}$ Faculty of Physics, Sofia University, Sofia, Bulgaria\\
$^{37}$ Faculty of Science, P.J.~\v{S}af\'{a}rik University, Ko\v{s}ice, Slovak Republic\\
$^{38}$ Frankfurt Institute for Advanced Studies, Johann Wolfgang Goethe-Universit\"{a}t Frankfurt, Frankfurt, Germany\\
$^{39}$ Fudan University, Shanghai, China\\
$^{40}$ Gangneung-Wonju National University, Gangneung, Republic of Korea\\
$^{41}$ Gauhati University, Department of Physics, Guwahati, India\\
$^{42}$ Helmholtz-Institut f\"{u}r Strahlen- und Kernphysik, Rheinische Friedrich-Wilhelms-Universit\"{a}t Bonn, Bonn, Germany\\
$^{43}$ Helsinki Institute of Physics (HIP), Helsinki, Finland\\
$^{44}$ High Energy Physics Group,  Universidad Aut\'{o}noma de Puebla, Puebla, Mexico\\
$^{45}$ Horia Hulubei National Institute of Physics and Nuclear Engineering, Bucharest, Romania\\
$^{46}$ Indian Institute of Technology Bombay (IIT), Mumbai, India\\
$^{47}$ Indian Institute of Technology Indore, Indore, India\\
$^{48}$ INFN, Laboratori Nazionali di Frascati, Frascati, Italy\\
$^{49}$ INFN, Sezione di Bari, Bari, Italy\\
$^{50}$ INFN, Sezione di Bologna, Bologna, Italy\\
$^{51}$ INFN, Sezione di Cagliari, Cagliari, Italy\\
$^{52}$ INFN, Sezione di Catania, Catania, Italy\\
$^{53}$ INFN, Sezione di Padova, Padova, Italy\\
$^{54}$ INFN, Sezione di Pavia, Pavia, Italy\\
$^{55}$ INFN, Sezione di Torino, Turin, Italy\\
$^{56}$ INFN, Sezione di Trieste, Trieste, Italy\\
$^{57}$ Inha University, Incheon, Republic of Korea\\
$^{58}$ Institute for Gravitational and Subatomic Physics (GRASP), Utrecht University/Nikhef, Utrecht, Netherlands\\
$^{59}$ Institute of Experimental Physics, Slovak Academy of Sciences, Ko\v{s}ice, Slovak Republic\\
$^{60}$ Institute of Physics, Homi Bhabha National Institute, Bhubaneswar, India\\
$^{61}$ Institute of Physics of the Czech Academy of Sciences, Prague, Czech Republic\\
$^{62}$ Institute of Space Science (ISS), Bucharest, Romania\\
$^{63}$ Institut f\"{u}r Kernphysik, Johann Wolfgang Goethe-Universit\"{a}t Frankfurt, Frankfurt, Germany\\
$^{64}$ Instituto de Ciencias Nucleares, Universidad Nacional Aut\'{o}noma de M\'{e}xico, Mexico City, Mexico\\
$^{65}$ Instituto de F\'{i}sica, Universidade Federal do Rio Grande do Sul (UFRGS), Porto Alegre, Brazil\\
$^{66}$ Instituto de F\'{\i}sica, Universidad Nacional Aut\'{o}noma de M\'{e}xico, Mexico City, Mexico\\
$^{67}$ iThemba LABS, National Research Foundation, Somerset West, South Africa\\
$^{68}$ Jeonbuk National University, Jeonju, Republic of Korea\\
$^{69}$ Johann-Wolfgang-Goethe Universit\"{a}t Frankfurt Institut f\"{u}r Informatik, Fachbereich Informatik und Mathematik, Frankfurt, Germany\\
$^{70}$ Korea Institute of Science and Technology Information, Daejeon, Republic of Korea\\
$^{71}$ KTO Karatay University, Konya, Turkey\\
$^{72}$ Laboratoire de Physique des 2 Infinis, Ir\`{e}ne Joliot-Curie, Orsay, France\\
$^{73}$ Laboratoire de Physique Subatomique et de Cosmologie, Universit\'{e} Grenoble-Alpes, CNRS-IN2P3, Grenoble, France\\
$^{74}$ Lawrence Berkeley National Laboratory, Berkeley, California, United States\\
$^{75}$ Lund University Department of Physics, Division of Particle Physics, Lund, Sweden\\
$^{76}$ Nagasaki Institute of Applied Science, Nagasaki, Japan\\
$^{77}$ Nara Women{'}s University (NWU), Nara, Japan\\
$^{78}$ National and Kapodistrian University of Athens, School of Science, Department of Physics , Athens, Greece\\
$^{79}$ National Centre for Nuclear Research, Warsaw, Poland\\
$^{80}$ National Institute of Science Education and Research, Homi Bhabha National Institute, Jatni, India\\
$^{81}$ National Nuclear Research Center, Baku, Azerbaijan\\
$^{82}$ National Research and Innovation Agency - BRIN, Jakarta, Indonesia\\
$^{83}$ Niels Bohr Institute, University of Copenhagen, Copenhagen, Denmark\\
$^{84}$ Nikhef, National institute for subatomic physics, Amsterdam, Netherlands\\
$^{85}$ Nuclear Physics Group, STFC Daresbury Laboratory, Daresbury, United Kingdom\\
$^{86}$ Nuclear Physics Institute of the Czech Academy of Sciences, Husinec-\v{R}e\v{z}, Czech Republic\\
$^{87}$ Oak Ridge National Laboratory, Oak Ridge, Tennessee, United States\\
$^{88}$ Ohio State University, Columbus, Ohio, United States\\
$^{89}$ Physics department, Faculty of science, University of Zagreb, Zagreb, Croatia\\
$^{90}$ Physics Department, Panjab University, Chandigarh, India\\
$^{91}$ Physics Department, University of Jammu, Jammu, India\\
$^{92}$ Physics Program and International Institute for Sustainability with Knotted Chiral Meta Matter (SKCM2), Hiroshima University, Hiroshima, Japan\\
$^{93}$ Physikalisches Institut, Eberhard-Karls-Universit\"{a}t T\"{u}bingen, T\"{u}bingen, Germany\\
$^{94}$ Physikalisches Institut, Ruprecht-Karls-Universit\"{a}t Heidelberg, Heidelberg, Germany\\
$^{95}$ Physik Department, Technische Universit\"{a}t M\"{u}nchen, Munich, Germany\\
$^{96}$ Politecnico di Bari and Sezione INFN, Bari, Italy\\
$^{97}$ Research Division and ExtreMe Matter Institute EMMI, GSI Helmholtzzentrum f\"ur Schwerionenforschung GmbH, Darmstadt, Germany\\
$^{98}$ Saga University, Saga, Japan\\
$^{99}$ Saha Institute of Nuclear Physics, Homi Bhabha National Institute, Kolkata, India\\
$^{100}$ School of Physics and Astronomy, University of Birmingham, Birmingham, United Kingdom\\
$^{101}$ Secci\'{o}n F\'{\i}sica, Departamento de Ciencias, Pontificia Universidad Cat\'{o}lica del Per\'{u}, Lima, Peru\\
$^{102}$ Stefan Meyer Institut f\"{u}r Subatomare Physik (SMI), Vienna, Austria\\
$^{103}$ SUBATECH, IMT Atlantique, Nantes Universit\'{e}, CNRS-IN2P3, Nantes, France\\
$^{104}$ Sungkyunkwan University, Suwon City, Republic of Korea\\
$^{105}$ Suranaree University of Technology, Nakhon Ratchasima, Thailand\\
$^{106}$ Technical University of Ko\v{s}ice, Ko\v{s}ice, Slovak Republic\\
$^{107}$ The Henryk Niewodniczanski Institute of Nuclear Physics, Polish Academy of Sciences, Cracow, Poland\\
$^{108}$ The University of Texas at Austin, Austin, Texas, United States\\
$^{109}$ Universidad Aut\'{o}noma de Sinaloa, Culiac\'{a}n, Mexico\\
$^{110}$ Universidade de S\~{a}o Paulo (USP), S\~{a}o Paulo, Brazil\\
$^{111}$ Universidade Estadual de Campinas (UNICAMP), Campinas, Brazil\\
$^{112}$ Universidade Federal do ABC, Santo Andre, Brazil\\
$^{113}$ University of Cape Town, Cape Town, South Africa\\
$^{114}$ University of Houston, Houston, Texas, United States\\
$^{115}$ University of Jyv\"{a}skyl\"{a}, Jyv\"{a}skyl\"{a}, Finland\\
$^{116}$ University of Kansas, Lawrence, Kansas, United States\\
$^{117}$ University of Liverpool, Liverpool, United Kingdom\\
$^{118}$ University of Science and Technology of China, Hefei, China\\
$^{119}$ University of South-Eastern Norway, Kongsberg, Norway\\
$^{120}$ University of Tennessee, Knoxville, Tennessee, United States\\
$^{121}$ University of the Witwatersrand, Johannesburg, South Africa\\
$^{122}$ University of Tokyo, Tokyo, Japan\\
$^{123}$ University of Tsukuba, Tsukuba, Japan\\
$^{124}$ University Politehnica of Bucharest, Bucharest, Romania\\
$^{125}$ Universit\'{e} Clermont Auvergne, CNRS/IN2P3, LPC, Clermont-Ferrand, France\\
$^{126}$ Universit\'{e} de Lyon, CNRS/IN2P3, Institut de Physique des 2 Infinis de Lyon, Lyon, France\\
$^{127}$ Universit\'{e} de Strasbourg, CNRS, IPHC UMR 7178, F-67000 Strasbourg, France, Strasbourg, France\\
$^{128}$ Universit\'{e} Paris-Saclay Centre d'Etudes de Saclay (CEA), IRFU, D\'{e}partment de Physique Nucl\'{e}aire (DPhN), Saclay, France\\
$^{129}$ Universit\`{a} degli Studi di Foggia, Foggia, Italy\\
$^{130}$ Universit\`{a} del Piemonte Orientale, Vercelli, Italy\\
$^{131}$ Universit\`{a} di Brescia, Brescia, Italy\\
$^{132}$ Variable Energy Cyclotron Centre, Homi Bhabha National Institute, Kolkata, India\\
$^{133}$ Warsaw University of Technology, Warsaw, Poland\\
$^{134}$ Wayne State University, Detroit, Michigan, United States\\
$^{135}$ Westf\"{a}lische Wilhelms-Universit\"{a}t M\"{u}nster, Institut f\"{u}r Kernphysik, M\"{u}nster, Germany\\
$^{136}$ Wigner Research Centre for Physics, Budapest, Hungary\\
$^{137}$ Yale University, New Haven, Connecticut, United States\\
$^{138}$ Yonsei University, Seoul, Republic of Korea\\
$^{139}$  Zentrum  f\"{u}r Technologie und Transfer (ZTT), Worms, Germany\\
$^{140}$ Affiliated with an institute covered by a cooperation agreement with CERN\\
$^{141}$ Affiliated with an international laboratory covered by a cooperation agreement with CERN.\\

\end{flushleft} 

\end{document}